\pdfoutput=1

\documentclass[useAMS,usenatbib,fleqn]{mnras}

\usepackage{graphicx}
\usepackage{amsmath,amssymb}

\title[Escape dynamics and NHIMs in star clusters]{Unraveling the escape dynamics and the nature of the normally hyperbolic invariant manifolds in tidally limited star clusters}

\author[E. E. Zotos \& Ch. Jung]{Euaggelos E. Zotos$^1$\thanks{E-mail: evzotos@physics.auth.gr} and Christof Jung$^2$\thanks{E-mail:jung@fis.unam.mx} \\
$^1$ Department of Physics, School of Science, Aristotle University of Thessaloniki, 541 24, Thessaloniki, Greece \\
$^2$ Instituto de Ciencias F\'{i}sicas, Universidad Nacional Aut\'{o}noma de M\'{e}xico Av. Universidad s/n, 62251 Cuernavaca, Mexico
}

\begin{document}

\date{Accepted 2016 September 28. Received 2016 September 28; in original form 2016 August 30}

\pubyear{2017} \volume{465} \pagerange{525--546}

\setcounter{page}{525}

\maketitle

\label{firstpage}

\begin{abstract}
The escape mechanism of orbits in a star cluster rotating around its parent galaxy in a circular orbit is investigated. A three degrees of freedom model is used for describing the dynamical properties of the Hamiltonian system. The gravitational field of the star cluster is represented by a smooth and spherically symmetric Plummer potential. We distinguish between ordered and chaotic orbits as well as between trapped and escaping orbits, considering only unbounded motion for several energy levels. The Smaller Alignment Index (SALI) method is used for determining the regular or chaotic nature of the orbits. The basins of escape are located and they are also correlated with the corresponding escape time of the orbits. Areas of bounded regular or chaotic motion and basins of escape were found to coexist in the $(x,z)$ plane. The properties of the normally hyperbolic invariant manifolds (NHIMs), located in the vicinity of the index-1 Lagrange points $L_1$ and $L_2$, are also explored. These manifolds are of paramount importance as they control the flow of stars over the saddle points, while they also trigger the formation of tidal tails observed in star clusters. Bifurcation diagrams of the Lyapunov periodic orbits as well as restrictions of the Poincar\'e map to the NHIMs are deployed for elucidating the dynamics in the neighbourhood of the saddle points. The extended tidal tails, or tidal arms, formed by stars with low velocity which escape through the Lagrange points are monitored. The numerical results of this work are also compared with previous related work.
\end{abstract}


\section{Introduction}
\label{intro}

Stars are born and developed from large clouds of molecular gas. These molecular clouds usually contain hundreds, or even thousands, of solar masses of material, so the stars form in groups of clusters. After the remnant gas is heated and blown away, the stars group together by their mutual gravitational attraction thus forming open star clusters.

Dynamical interactions between the stars are the main reason why star clusters have the tendency to gradually loose mass until they completely dissolved. A substantial amount of stars of the clusters become members of the disc and the halo populations of the parent galaxy, as the clusters dissolve over time. The dissolution mechanism of star clusters is a very active as well as challenging field of research. Theoretical aspects of this intricate subject, like the evaporation process because of the movement of stars above the escape velocity and the time-scale of relaxation which determines the rate of the dynamical evolution of a star cluster were the first which have been investigated analytically \citep[e.g.,][]{A38,S40}. Later on, the dynamical properties of stars which have been scattered above the critical escape energy were examined \citep{K59}, while a year later the importance of the close encounters between stars for the rate of mass loss of a star cluster was studied \citep{H60}.

According to \citet{LL03} a considerable amount of the stars of a galaxy are actually born in star clusters. In particular, some recent studies shed some light on the mechanism of how stars are in fact born and develop star clusters \citep[e.g.,][]{BBG10,K12}. Inevitably, all star clusters dissolve over time due to two main reasons: (i) the two-body process (encounters) that affects all the members of a cluster thus forcing them to obtain escape velocities and (ii) the strong tidal forces due to the external gravitational attraction of the parent galaxy. More precisely, the tidal field of the parent galaxy can significantly affect and therefore disturb the orbital behaviour of the star cluster itself \citep{RMH97}.

Usually we use the term ``tidally limited" star cluster which means, roughly speaking, that the tidal field imposes a boundary, outside of which the mass density vanishes, while the star cluster is being captured within the tidal limit. During the energy exchange between the stars, some of them reach the escape velocity and become runaway stars. The escape of stars form a tidally limited cluster is a two-stage process. In the first stage, the stars are scattered into the escaping phase space by two-body encounters, while at the second stage they escape through the exit channels in the open equipotential surface. The time required for a star to complete the first stage strongly depends on the relaxation time, while on the other hand, the time needed for a star to complete stage two is mainly related to the energy the star has. In \citet{FH00} an expression regarding the escape time of stars in the second stage which have just finished stage one had been derived, while in \citet{B01} the corresponding results had been exploited in order to address the important issue of the dissolution time of star clusters. Furthermore, some additional interesting aspects, like the velocity of escaping stars \citep[e.g.,][]{KG07,SN11,LLZ12}, or their population type \citep{R97} have also been studied.

Complicated stellar formations, such as tidal tails or tidal arms, are formed by stars which escape from the cluster and they are captured by the strong gravitational field of the parent galaxy. These interesting stellar formations have been observed in the Milky Way \citep[e.g.,][]{GFI95,KSL97,LS97,OGR01,ROG02,LLF03,BEI06} and they have also been modelled and explored \citep[e.g.,][]{JSH99,YL02,DOG04,KGO04,CMM05,dMCM05,LLS06,CWK07,FEB07,MCd07,KMH08,JBPE09,KKBH10}. At this point we should emphasize, that all the above-mentioned references on tidal tails and tidal arms are exemplary rather than exhaustive.

In \citet{EJSP08} (hereafter Paper I) the escape dynamics of a 2-dof model of a star cluster embedded in the tidal field of a parent galaxy was investigated, while in \citet{Z15a} (hereafter Paper II) the exploration was expanded into three dimensions by using a 3-dof model. In both papers the orbital dynamics and the corresponding basins of escape were determined and obtained by conducting a thorough and systematic orbit classification in the available phase space. In \citet{Z15b} the escape dynamics of a tidally limited star cluster was compared by using different types of spherically symmetric gravitational potentials for describing the properties of the star cluster.

If the energy is above, but close, to the saddle energy of the relevant saddle of the effective potential then the escape process is directed and channeled by the stable and unstable manifolds of some normally hyperbolic invariant manifolds (NHIMs) sitting over this saddle. On this basis, a central part of the work of this article will consist in the study of the NHIMs. The Lyapunov orbits \citep{L07} are important parts of the NHIMs. Therefore, also the development scenario of the Lyapunov orbits will be investigated in detail. We apply recently developed methods \citep[e.g.,][]{GDJ14,GJ15} for the numerical study of NHIMs and of the Poincar\'e map restricted to the NHIM. This restricted map acts on a 2-dimensional domain, can be represented by 2-dimensional graphics and therefore it is an ideal tool to present graphically the development scenario of codimension 2 NHIMs in 3-dof systems. In a recent paper \citep{JZ16} (hereafter Paper III) the development scenario of codimension 2 NHIMs in a 3-dof model of a barred galaxy has been numerically investigated.

The structure of the present paper is as follows: In Section \ref{mod} we briefly describe the main properties of the tidally limited star cluster model. In the following Section, we numerically investigate the escape dynamics of stars. Section \ref{nhims} contains a detailed description of the dynamics in the neighbourhood of the saddle points and in particular of the NHIMs sitting near these points. In Section \ref{tdt} we link the invariant manifolds with the tidal tail structures observed in star clusters. Our paper continues with Section \ref{conc}, where the main conclusions of our work are presented. The paper ends with two appendices, in the first one we explain the connection between the total potential and the effective potential, while in the second one we illustrate the scaling behaviour near the stable manifolds of NHIMs.

\section{Properties of the dynamical model}
\label{mod}

In order to describe the dynamical properties of a star cluster we need to use the so-called ``tidal approximation" model. According to this theory we assume that the star cluster revolves around the center of a parent galaxy on a circular orbit and with constant angular velocity $\Omega$. Using this assumption we are allowed to apply the epicyclic approximation for determining the tidal forces which act on all stars of the cluster. For this purpose, the most appropriate system of coordinates is an accelerated rotating frame of reference \citep{C42}. In this system the center of the parent galaxy as well as the star cluster itself are at rest. Moreover, the origin of the coordinates is located at the center of the star cluster. This assures that the position of the galactic center is $P(x,y,z) = \left(-R_{\rm g}, 0, 0\right)$, where of course $R_{\rm g}$ is the radius of the circular orbit of the star cluster.

Taking into account that the star cluster is spherically symmetric the best choice for describing its properties is by using the simple, yet self-consistent Plummer potential (more details about the isolated Plummer potential can be found in the Appendix A of Paper I)
\begin{equation}
\Phi_{\rm cl}(x,y,z) = - \frac{G M_{\rm cl}}{\sqrt{x^2 + y^2 + z^2 + r_{\rm Pl}^2}},
\label{potcl}
\end{equation}
where $G$ is the gravitational constant, while $M_{\rm cl}$ is the total mass of the star cluster. The Plummer radius is $r_{\rm Pl} = 0.182$ and it was chosen in such a way that the potential (\ref{potcl}) is the best fit to a King model with $W_0 = 4$, where $W_0$ is a parameter related to the mass density of the cluster \citep[see e.g.,][]{K66}.

Considering the relationships connecting the tidal forces with the epicyclic frequency $\kappa$ as well as with the vertical frequency $\nu$ \citep[see for more details][]{BT08} and also with the rotation of the star cluster we conclude that the total potential\footnote{We would like to clarify that in Paper I the effective potential $\Phi_{\rm eff}$ was presented, while now the total potential $\Phi_{\rm t} = \Phi_{\rm eff} + \frac{1}{2}\left(x^2 + y^2\right)$ is used. At the end of the paper, in the Appendix A, we shall explain in detail how these two different concepts and their different functional forms are related.} is
\begin{align}
\Phi_{\rm t}(x,y,z) &= \Phi_{\rm cl}(x,y,z) + \frac{1}{2}\left(\kappa^2 - 4\Omega^2 \right) x^2 + \frac{1}{2}\nu^2 z^2 \nonumber \\
&+ \frac{\Omega^2}{2}\left(x^2 + y^2\right).
\label{vt}
\end{align}

Then the corresponding equations of motion read
\begin{align}
\dot{x} &= p_x + \Omega y, \nonumber \\
\dot{y} &= p_y - \Omega x, \nonumber \\
\dot{z} &= p_z, \nonumber \\
\dot{p_x} &= - \frac{\partial \Phi_{\rm t}}{\partial x} + \Omega p_y, \nonumber \\
\dot{p_y} &= - \frac{\partial \Phi_{\rm t}}{\partial y} - \Omega p_x, \nonumber \\
\dot{p_z} &= - \frac{\partial \Phi_{\rm t}}{\partial z},
\label{eqmot}
\end{align}
where, as usual, the dot indicates the derivative with respect to the time, while $p_x$, $p_y$ and $p_z$ are the canonical momenta per unit mass, conjugate to $x$, $y$ and $z$, respectively.

For the computation of standard chaos indicators (the SALI in this case) we need to follow the time-evolution of deviation vectors ${\bf{w_i}}, i = 1,2$. Therefore the required set of variational equations is
\begin{align}
\dot{(\delta x)} &= \delta p_x + \Omega \delta y, \nonumber \\
\dot{(\delta y)} &= \delta p_y - \Omega \delta x, \nonumber \\
\dot{(\delta z)} &= \delta p_z, \nonumber \\
(\dot{\delta p_x}) &=
- \frac{\partial^2 \Phi_{\rm t}}{\partial x^2} \ \delta x
- \frac{\partial^2 \Phi_{\rm t}}{\partial x \partial y} \delta y
- \frac{\partial^2 \Phi_{\rm t}}{\partial x \partial z} \delta z + \Omega \delta p_y, \nonumber \\
(\dot{\delta p_y}) &=
- \frac{\partial^2 \Phi_{\rm t}}{\partial y \partial x} \delta x
- \frac{\partial^2 \Phi_{\rm t}}{\partial y^2} \delta y
- \frac{\partial^2 \Phi_{\rm t}}{\partial y \partial z} \delta z - \Omega \delta p_x, \nonumber \\
(\dot{\delta p_z}) &=
- \frac{\partial^2 \Phi_{\rm t}}{\partial z \partial x} \delta x
- \frac{\partial^2 \Phi_{\rm t}}{\partial z \partial y} \delta y
- \frac{\partial^2 \Phi_{\rm t}}{\partial z^2} \delta z.
\label{vareq}
\end{align}

The equations of motion (\ref{eqmot}) admit the following isolating energy integral, which governs the motion of a test particle (star) with a unit mass $(m = 1)$
\begin{equation}
H = \frac{1}{2} \left(p_x^2 + p_y^2 + p_z^2 \right) + \Phi_{\rm t}(x,y,z) - \Omega L_z = E,
\label{ham}
\end{equation}
where $E$ is the numerical value of the energy integral which is conserved, while $L_z = x p_y - y p_x$ is the angular momentum along the $z$ direction. At this point, it should be emphasized that if the star cluster follows another type of orbit (i.e., elliptical instead of circular) around the parent galaxy, all numerical calculations become much more difficult to be performed because no integral of motion, comparable to the energy integral, is known.

The whole system and the equations of motion of Eq. (\ref{eqmot}) have two discrete symmetries. First, they are invariant under the transformation $z \to -z$, $p_z \to - p_z$, i.e. under a reflection in the symmetry plane $z = 0$. Second, the more interesting symmetry is a simultaneous inversion of all 4 horizontal coordinates, i.e. the transformation $x \to -x$, $y \to -y$, $p_x \to - p_x$, and $p_y \to -p_y$. This transformation is equivalent to a rotation of the whole system around the $z$-axis by an angle $\pi$.

This second symmetry is a consequence of the tidal approximation. As Eq. (\ref{vt}) shows some quadratic approximation has been made for the gravitational potential of the parent galaxy and the approximated potential contains even terms only. This property is the source of the symmetry. Any inclusion of odd terms into the total potential of Eq. (\ref{vt}) would destroy this symmetry. It is also obvious that any two symmetry related orbits have the same value of the energy because the Hamiltonian of the system is invariant under the symmetry operation. Some consequences of this symmetry will be discussed
in later sections and it will be referred to as the inversion symmetry of the system.

For all the involved dynamical quantities we use the dimensionless system of units introduced in Paper I according to which $G = 1$, $\Omega = 1$ and $M_{\rm cl} = 4 \Omega^2 - \kappa^2 = 2.2$. Furthermore, the unit of length is one tidal radius $r_{\rm t} = 1$. The characteristic frequencies $\Omega$, $\kappa$ and $\nu$ on the other hand, are related to the galactic gravitational potential and in the neighborhood of our Sun can be expressed as functions of the Oort's constants \citep[more details can be found in][]{BT08}. Using the numerical values of the Oort's constants provided in \citet{FW97}, we obtain the following values: $\kappa^2/\Omega^2 \simeq 1.8$ and $\nu^2/\Omega^2 \simeq 7.6$.

A fundamental scale length of the star cluster is the tidal radius \citep{K62} which is defined as
\begin{equation}
r_{\rm t} = \left(\frac{G M_{\rm cl}}{4\Omega^2 - \kappa^2}\right)^{1/3}.
\label{rt}
\end{equation}

\begin{figure*}
\centering
\resizebox{\hsize}{!}{\includegraphics{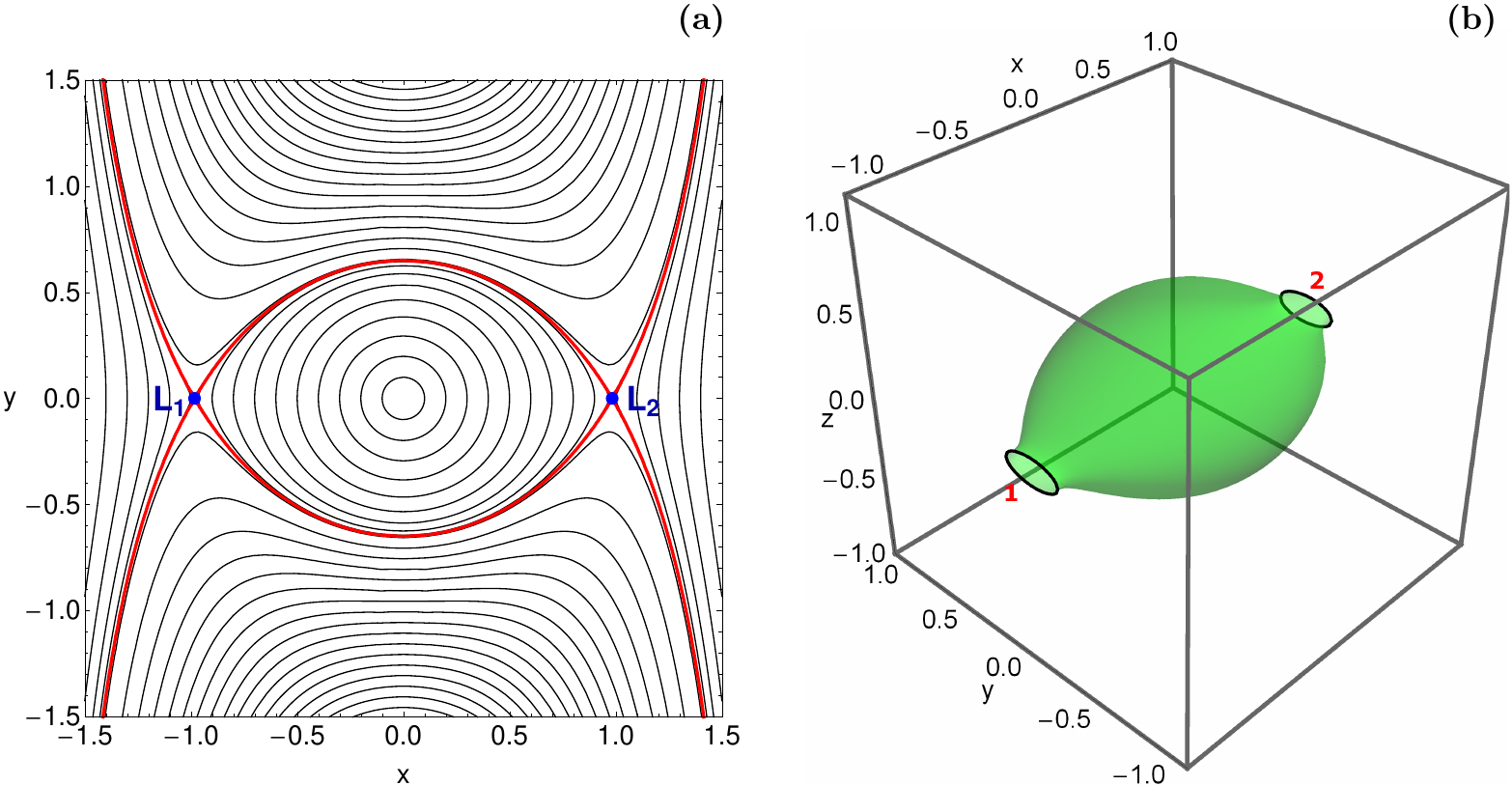}}
\caption{(a-left): The isoline contours of the effective potential $\Phi_{\rm eff}$ of the star cluster on the $(x,y)$-plane for $z = 0$. The equipotential curve corresponding to the critical energy level $E_L$ is shown in red, while the position of the two Lagrange points $L_1$ and $L_2$ are indicated by blue dots. (b-right): The three-dimensional equipotential surface of the effective potential $\Phi_{\rm eff}(x,y,z) = E$ when $E = -3.23$. The stars of the cluster can leak out through the exit channels 1 and 2 thus passing either through $L_1$ or $L_2$, respectively. (For the interpretation of references to colour in this figure caption and the corresponding text, the reader is referred to the electronic version of the article.)}
\label{pot}
\end{figure*}

The effective potential $\Phi_{\rm eff}$ has two Lagrange saddle points $L_1$ and $L_2$ which are located at $L_1(x,y,z) = (-r_{\rm L}, 0, 0)$ and $L_2(x,y,z) = (r_{\rm L}, 0, 0)$, respectively where $r_{\rm L} = 0.98329853$ is the Lagrange radius\footnote{It is interesting to note that the Lagrange radius is almost equal to the tidal radius.}. The numerical value of the effective potential at the Lagrange points yields a critical energy level $E_L = -3.2635636$ (the corresponding equipotential curve is shown in red in Fig. \ref{pot}a) which is of course the energy of escape. When $E = E_L$ the equipotential surface encloses the critical volume, while for $E > E_L$ the equipotential surface is open and consequently stars are free to escape from the cluster. In Fig. \ref{pot}b we present a plot of the three-dimensional equipotential surface $\Phi_{\rm eff}(x,y,z) = E$ when $E = -3.23 > E_L$. We observe the two openings (exit channels or throats) through which the test particles can leak out. In fact, we may say that these two channels around the Lagrange points act as hoses thus connecting the interior region $(-r_{\rm L} \leq x \leq r_{\rm L})$ of the cluster with the exterior region, or in other words the ``outside world". Exit channel 1 (negative $x$-direction) indicates escape towards the galactic center, while channel 2 (positive $x$-direction) indicates escape towards infinity.

A double precision Bulirsch-Stoer \verb!FORTRAN 77! algorithm \citep[e.g.,][]{PTVF92} was used for integrating backwards and forwards the equations of motion (\ref{eqmot}) as well as the variational equations (\ref{vareq}). The time step of the numerical integration was of the order of $10^{-2}$ which is sufficient enough for the desired accuracy of our computations. Throughout our calculations the numerical error regarding the conservation of the energy integral of motion of Eq. (\ref{ham}) was smaller than $10^{-13}$, although there were cases that the corresponding error was smaller than $10^{-14}$. All graphical illustrations presented in this paper have been created using version 10.3 of Mathematica$^{\circledR}$ \citep[e.g.,][]{Wolf03}.

\section{Escape dynamics}
\label{esc}

In Paper II we investigated for the first time the escape dynamics of the three degrees of freedom system of the star cluster. In particular, we explored the orbital structure of the configuration $(x,y)$ space as well as the phase $(x,\dot{x})$ space for several initial values of the $z$ coordinate. However it is not possible to fully understand the escape process of stars by considering only some specific values of $z_0$.

In this work in order to obtain a more complete view regarding the escape mechanism of stars in tidally limited star clusters we shall conduct some additional numerical calculations. It would be very illuminating to unveil how the initial value of $z$ influences the escape process of the orbits. To achieve this we decided to define, for several energy levels $E$, dense uniform grids of $1024 \times 1024$ initial conditions on the $(x,z)$ plane inside the corresponding energetically allowed area. Following a typical approach, all orbits of the stars are launched with initial conditions inside the tidal radius $(x_0^2 + y_0^2 + z_0^2 \leq r_{\rm t}^2)$. All three-dimensional orbits have initial conditions $(x_0,z_0)$ with $y_0 = p_{x0} = p_{z0} = 0$, while the initial value of $p_y$ is always obtained from the energy integral (\ref{ham}) as $p_{y_0} = p_y(x_0,y_0,z_0,p_{x_0},p_{z0},E)$ (we choose the branch of $p_y$ with positive orientation of intersection of the plane $y = 0$).

Our task will be to distinguish between bounded and unbounded (escaping) motion. In Paper II we found a substantial amount of trapped chaotic orbits, which do escape only after extremely long integration time. Therefore it would be very useful to classify initial conditions of bounded orbits into two types: (i) non-escaping regular orbits and (ii) trapped chaotic orbits. Over the years several dynamical indicators have been developed for distinguishing between order and chaos. As in Paper II we choose to use the Smaller ALingment Index (SALI) method \citep{S01} which has been proved a very fast yet reliable tool. The nature of an orbit can be determined by the numerical value of SALI at the end of the numerical integration. Being more precise, if SALI $> 10^{-4}$ the orbit is ordered, while if SALI $< 10^{-8}$ the orbit is chaotic. On the other hand, when the value of SALI lies in the interval $[10^{-8}, 10^{-4}]$ we have the case of a sticky orbit and further numerical integration is needed to fully reveal the true character of the orbit.

For the numerical integration of the initial conditions of the orbits we set a maximum time of $10^4$ time units. Our previous experience suggests that in most cases (energy levels) the majority of the orbits require considerable less time for finding one of the exit channels in the equipotential surface and therefore escape (obviously, the numerical integration is effectively stopped when an orbit escapes through the Lagrange points). Nevertheless, just for being sure that all orbits have enough time to escape, we decided to use such a long integration time. Thus, any orbit which remains trapped inside the tidal radius after an integration time of $10^4$ time units will be considered as a trapped (regular or chaotic) one.

In the following we shall classify initial conditions of orbits into four categories: (i) trapped chaotic orbits, (ii) non-escaping regular, (iii) escaping through $L_1$, and (iv) escaping through $L_2$. At the same time, we shall monitor the escape time (we will also use the terms escape rate and escape period) of the orbits.

In Paper II we performed a similar orbit classification however for energy levels relatively close to the escape energy. In this case, we will proceed to energy levels a lot higher than the energy of escape and specifically in the interval $E \in (E_L,-1]$. We restrict our analysis to this energy interval because if the energy is very high above the saddle $(E > -1)$, then this saddle is no longer important for the escape. The orbits just go out as they want into almost any direction and they are no longer restricted by the saddle structures, as for example the NHIM and its stable and unstable manifolds. In addition if we think from the point of view of the star cluster, then stars with a very high energy have left the cluster immediately after their formation and have gone a long time ago. Millions of years after the formation only such stars are left with an energy around or below the saddle energy. And from time to time one of those stars gains a little energy by direct interactions with other stars and comes a little above the saddle energy. For such stars our investigations become relevant.

\begin{figure*}
\centering
\resizebox{\hsize}{!}{\includegraphics{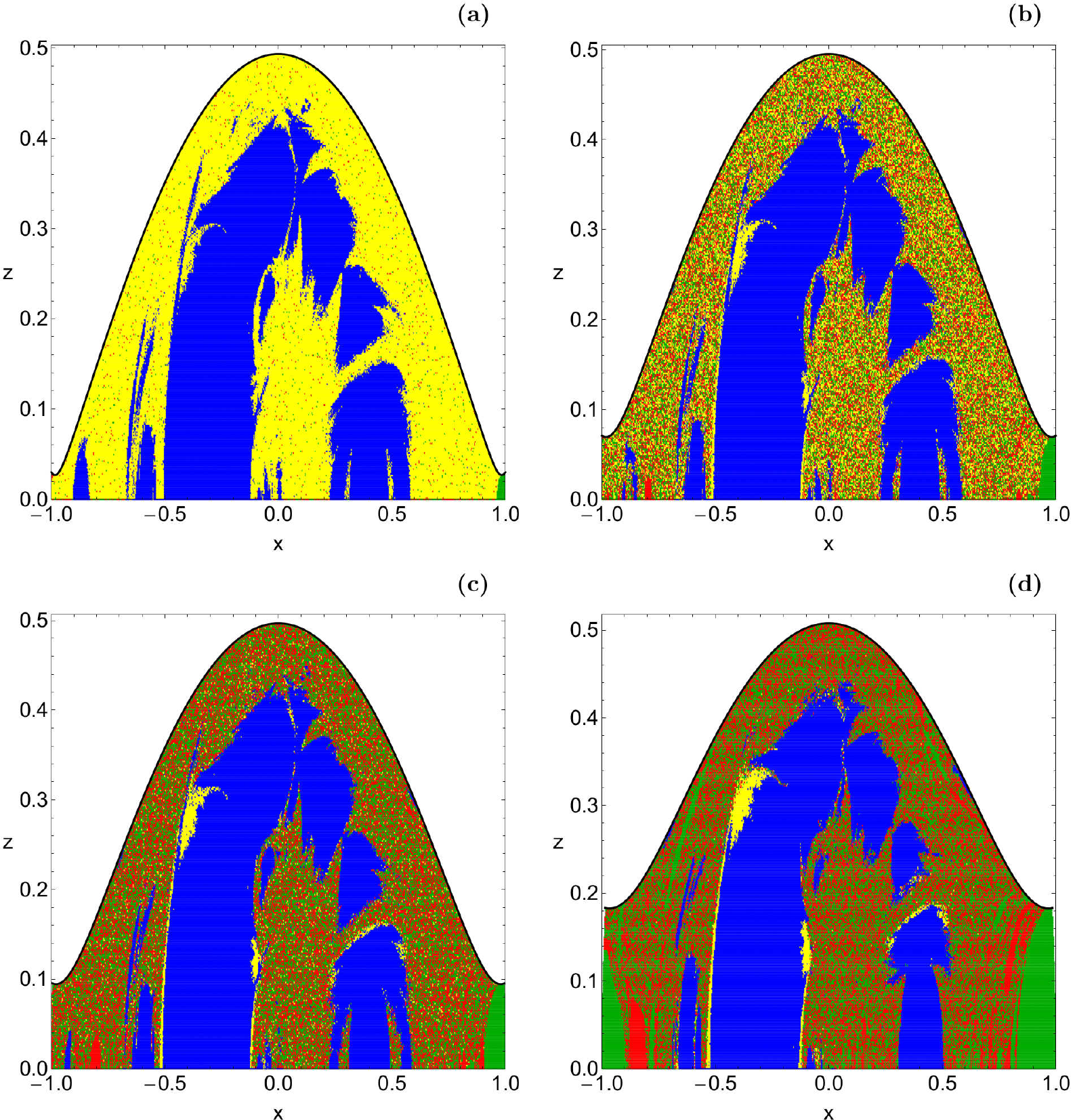}}
\caption{Orbital structure of the $(x,z)$ plane when (a-upper left): $E = -3.26$; (b-upper right): $E = -3.24$; (c-lower left): $E = -3.22$; (d-lower right): $E = -3.10$. The colour code is as follows: non-escaping regular orbits (blue), trapped chaotic orbits (yellow), escaping orbits through $L_1$ (red), orbits escaping orbits through $L_2$ (green). (For the interpretation of references to colour in this figure caption and the corresponding text, the reader is referred to the electronic version of the article.)}
\label{xz1}
\end{figure*}

\begin{figure*}
\centering
\resizebox{\hsize}{!}{\includegraphics{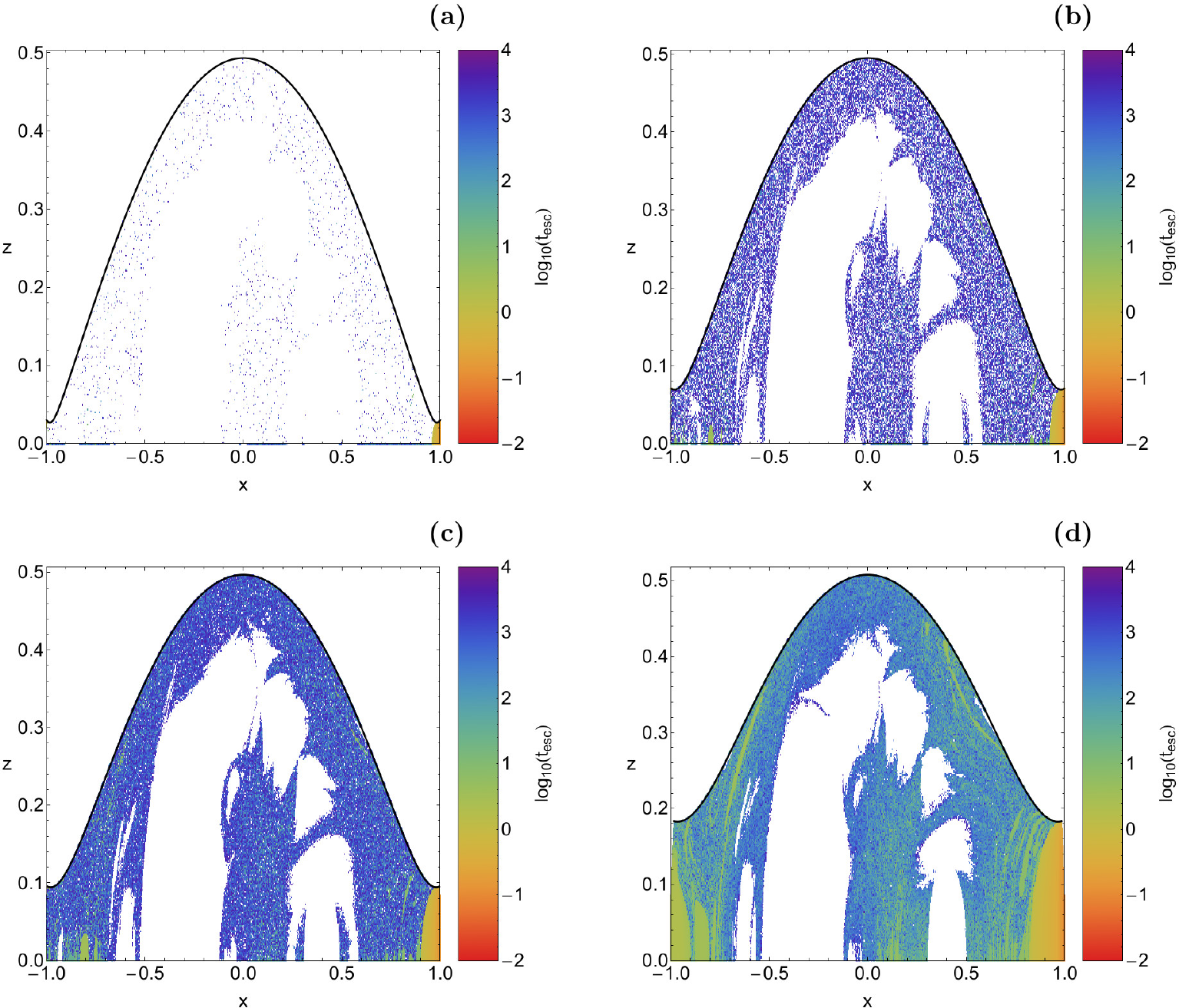}}
\caption{Distribution of the corresponding escape time $t_{\rm esc}$ of the orbits on the $(x,z)$ plane for the values of the energy presented in panels (a-d) of Fig. \ref{xz1}, respectively. The darker the colour, the higher the escape time. Initial conditions of non-escaping regular orbits and trapped chaotic orbits are shown in white. (For the interpretation of references to colour in this figure caption and the corresponding text, the reader is referred to the electronic version of the article.)}
\label{xzt1}
\end{figure*}

\begin{figure*}
\centering
\resizebox{\hsize}{!}{\includegraphics{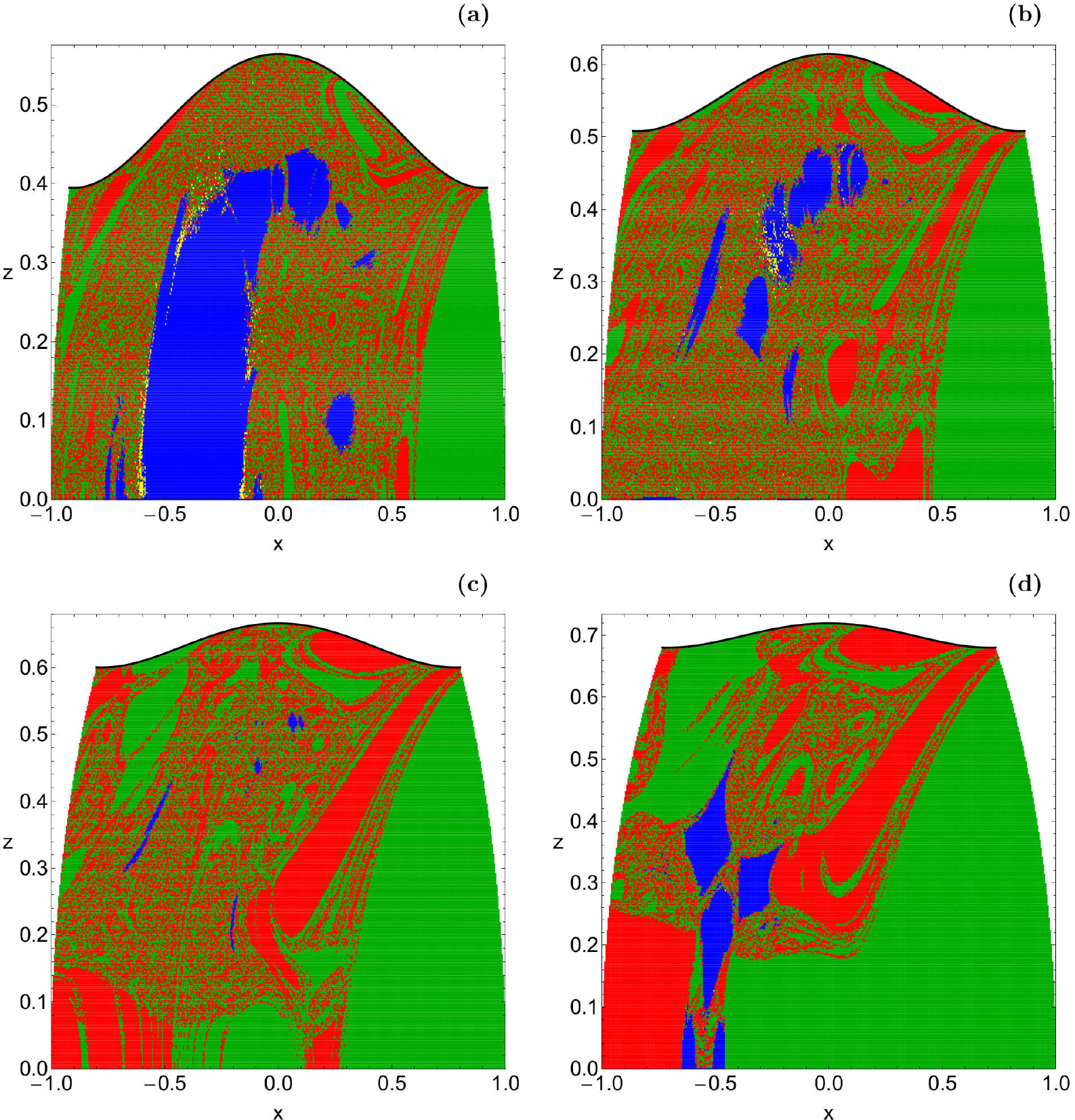}}
\caption{Orbital structure of the $(x,z)$ plane when (a-upper left): $E = -2.5$; (b-upper right): $E = -2.0$; (c-lower left): $E = -1.5$; (d-lower right): $E = -1.0$. The colour code is the same as in Fig. \ref{xz1}. (For the interpretation of references to colour in this figure caption and the corresponding text, the reader is referred to the electronic version of the article.)}
\label{xz2}
\end{figure*}

\begin{figure*}
\centering
\resizebox{\hsize}{!}{\includegraphics{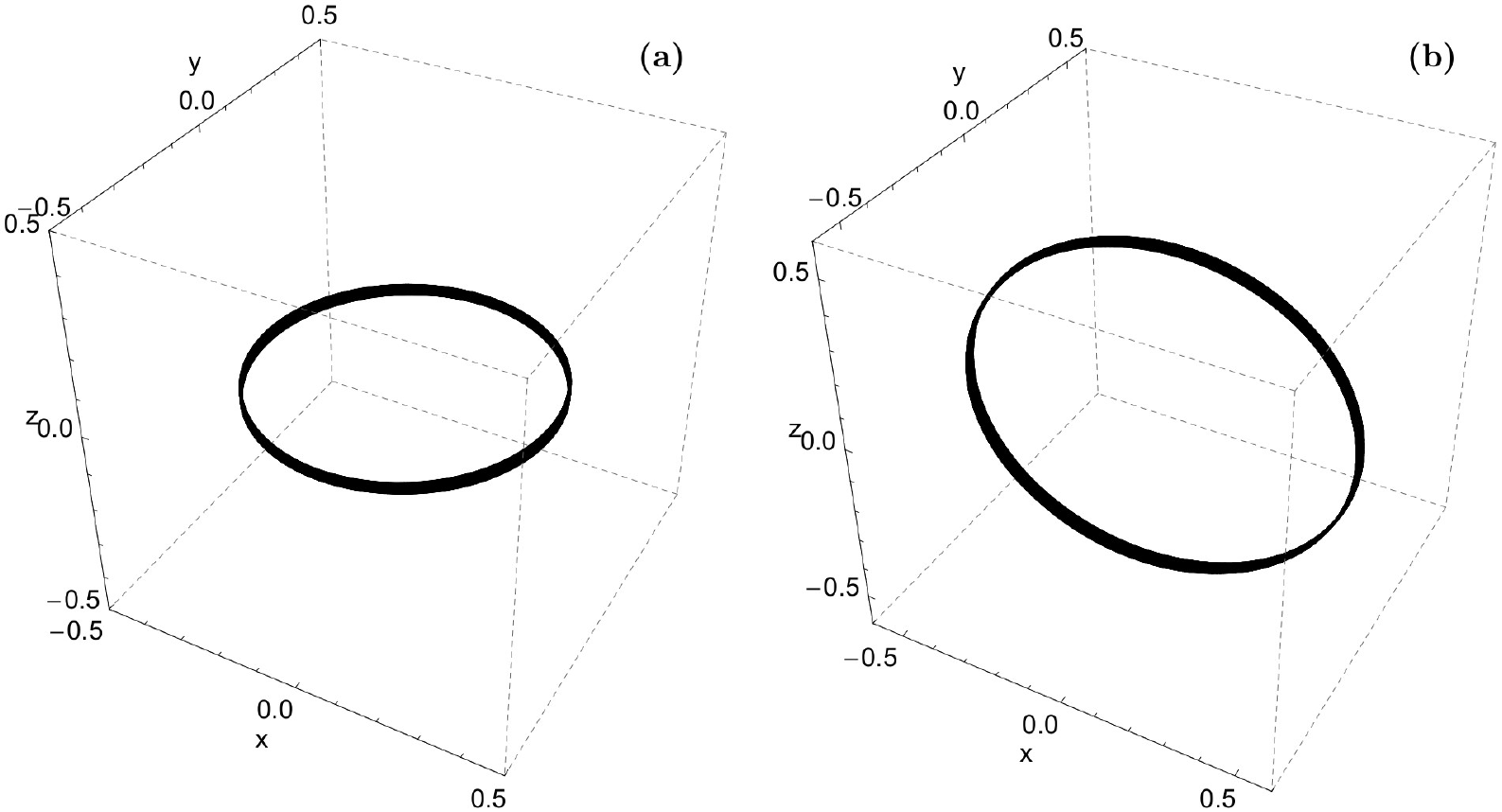}}
\caption{(a-left): A 3D 1:1:0 loop orbit parallel to the $(x,y)$ plane with initial conditions $x_0 = -0.36$, $z_0 = 0.015$, for $E = -2.5$. (b-right): A 3D 1:1:0 titled loop orbit with initial conditions $x_0 = -0.5$, $z_0 = 0.25$, for $E = -1.0$. In both cases $y_0 = p_{x0} = p_{z0} = 0$, while the initial value of $p_y$ is obtained for the energy integral (\ref{ham}).}
\label{orbs0}
\end{figure*}

\begin{figure*}
\centering
\resizebox{\hsize}{!}{\includegraphics{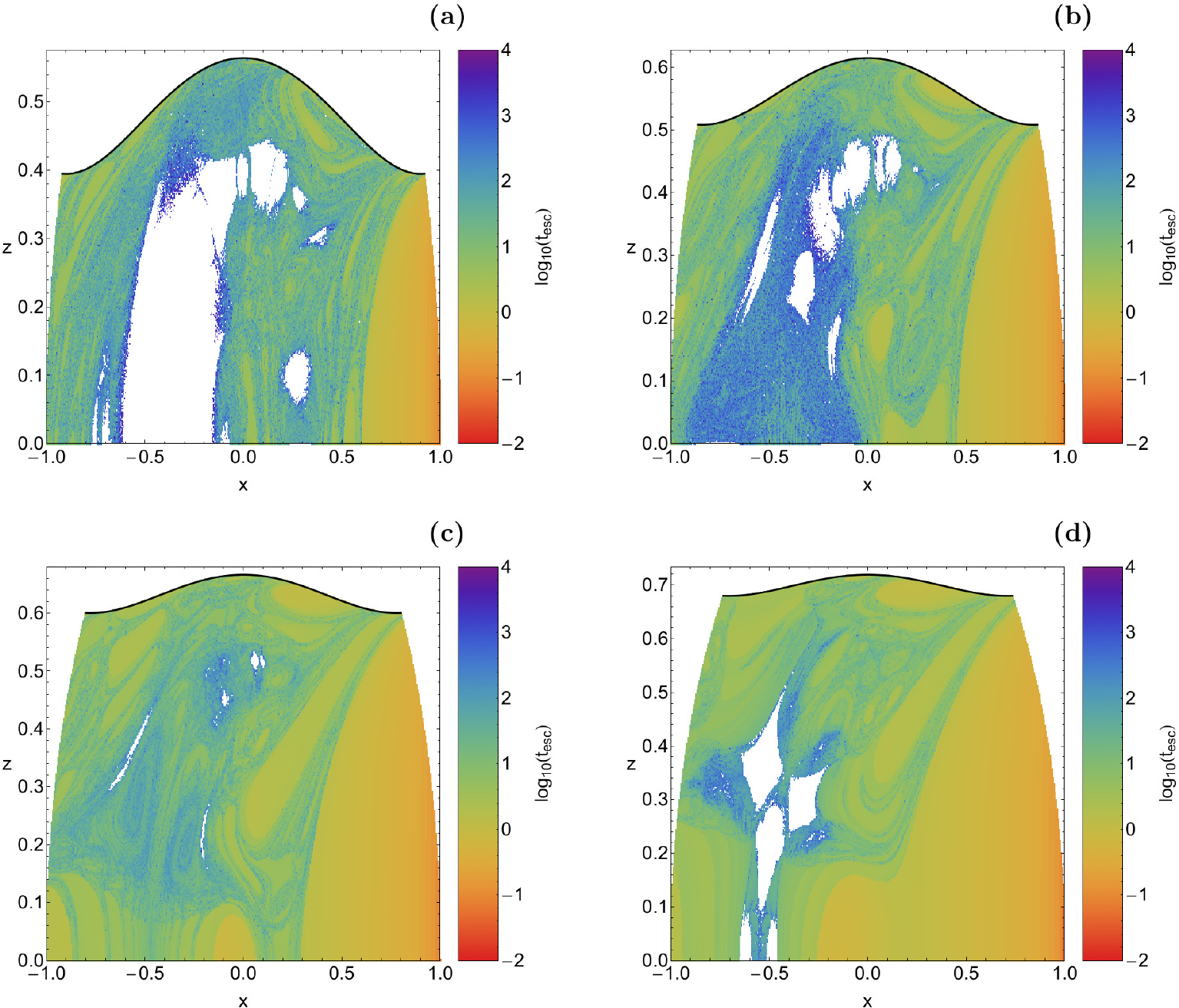}}
\caption{Distribution of the corresponding escape time $t_{\rm esc}$ of the orbits on the $(x,z)$ plane for the values of the energy presented in panels (a-d) of Fig. \ref{xz2}, respectively. The colour code is the same as in Fig. \ref{xzt1}. (For the interpretation of references to colour in this figure caption and the corresponding text, the reader is referred to the electronic version of the article.)}
\label{xzt2}
\end{figure*}

The orbital structure of the $(x,z)$ plane for four values of the energy is presented in Fig. \ref{xz1}(a-d). In these colour-coded grids we assign to each pixel a specific colour according to the corresponding type of the orbit. In particular, blue colour corresponds to regular non-escaping orbits, yellow colour corresponds to trapped chaotic orbits, red colour corresponds to orbits escaping through exit channel 1, while the initial conditions of orbits that escape through exit channel 2 are marked with green colour. The outermost black solid line denotes the zero velocity curve which is defined as
\begin{equation}
f(x,z) = \Phi_{\rm eff}(x, y = 0, z) = E.
\label{zvc}
\end{equation}

For $E = -3.26$, which is an energy level just above the energy of escape, we observe in Fig. \ref{xz1}a that more than half of the $(x,z)$ plane is covered by initial conditions of trapped chaotic orbits. The phenomenon of trapped chaos for energy levels above yet very close to the energy of escape has also been observed in Paper II. Inside the vast chaotic sea we can identify initial conditions of escaping orbits. However escape is rather difficult for such energy levels. A substantial amount of the $(x,y)$ plane is occupied by numerous islands corresponding to non-escaping regular motion. These stability islands correspond mainly to 1:1:0 resonant loop orbits \citep[see also][]{FH00}. As the value of the energy increases it is seen, in panels (b-d) of Fig. \ref{xz1}, that the portion of trapped chaotic orbits decreases, while at the same time, the amount of escaping orbits grows rapidly. On the other hand, the structures of the stability islands remains almost unperturbed. In panel (d), where $E = -3.10$, one may see that initial conditions of trapped chaotic orbits are mainly present in the vicinity of the boundaries of the stability islands. Now the majority of the $(x,z)$ plane is covered by a high fractal\footnote{We would like to emphasize that when it is stated that an area is fractal we simply mean that it has a fractal-like geometry without conducting any specific calculations as in \citet{AVS09}.} mixture of escaping orbits. In this domain a high dependence of the escape mechanism on the particular initial conditions of the orbits is observed. This means that a minor change in the $(x_0,z_0)$ initial conditions of the orbit has as a result the star to escape through the opposite escape channel which is of course a classical indication of chaotic motion. It is interesting to note that especially at the outer parts of the $(x,z)$ plane, near the Lagrange points, basins of escape\footnote{A local set of initial conditions that corresponds to a certain escape channel defines a basin of escape.} are present.

The corresponding distribution of the escape time $t_{\rm esc}$ of the orbits presented in Fig. \ref{xz1}(a-d) is illustrated in Fig. \ref{xzt1}(a-d). Once more, the scale of the escape time is revealed through a colour code. Specifically, initial conditions of fast escaping orbits are indicating with light reddish colors, while dark blue/purple colors suggest high escape periods. Furthermore, initial conditions of non-escaping and trapped chaotic obits are shown in white. Note that the colour bar has a logarithmic scale. We observe in panels (a) and (b), where $E = -3.26$ and $E = -3.24$, respectively that the escape time of the orbits are huge corresponding to several thousands of time units. However, as the value of the energy increases the escape rates of the orbits are constantly being reduced. This is true, because when $E = -3.10$ it is seen in panel (d) of Fig. \ref{xzt1} that the majority of the escaping orbits need about a couple of hundred time units in order to escape. Looking carefully the time distributions it becomes evident that orbits with initial conditions in the fractal areas of the plot require a significant amount of time before they escape from the cluster. On the contrary, the shortest escape rates have been measured for orbits inside the basins of escape where there is no dependence on the initial conditions whatsoever.

The escape dynamics of the dynamical system for four additional values of the energy is given in Fig. \ref{xz2}(a-d). Again different colours are used for distinguishing between the four types of the orbits (non-escaping regular, trapped chaotic, escaping through $L_1$ and escaping through $L_2$). In this case all four energy levels are much higher than the energy of escape. In Fig. \ref{xz2}a, where $E = -2.5$, we observe that escaping orbits cover about 80\% of the $(x,z)$ plane by forming well-defined basins of escape. The size of stability islands corresponding to non-escaping regular motion has been reduced and they are mainly confined to the $(x < 0)$ part of the $(x,z)$ plane. Therefore the majority of the ordered orbits are in fact retrograde 1:1:0\footnote{The $n:m:l$ notation we use for classifying the regular 3D orbits is according to \citet{CA98} and \citet{ZC13}. The ratio of those integers corresponds to the ratio of the main frequencies of the orbit, where the main frequency is the frequency of the greatest amplitude for each coordinate. Main amplitudes, when having a rational ratio, define the resonance of an orbit.} resonant orbits (i.e., when a star revolves around the star cluster in the opposite sense with respect to the motion of the cluster around the parent galaxy). Nevertheless, prograde 1:1:0 resonant orbits with $x > 0$ are also present. When $E = -2.0$ it is seen in panel (b) of Fig. \ref{xz2} that the stability regions are further reduced, while the main stability island which was present, so far, splits into many pieces. At the same time the extent of the basins of escape grows even further. In panel (c) of the same figure, where the energy is $E = -1.5$, one may see that the stability islands are hardly visible, while escape regions dominate the $(x,z)$ plane. In particular, basins of escape corresponding to escape through $L_2$ cover about 60\% of the grid. Our computations suggest that for $E > -2.0$ trapped chaotic motion is almost negligible as the initial conditions of such orbits appear only as lonely points either at the boundaries of the stability islands or randomly scattered in the vast escape region.

The highest energy level for which we have numerically investigated the orbital structure of the $(x,z)$ plane is $E = -1.0$. Our results are presented in panel (d) of Fig. \ref{xz2}. For this energy level we observe a very interesting phenomenon. Stability islands of non-escaping regular orbits emerge inside the vast escape region in the left side $(x < 0)$ of the plane. Taking into account the outcomes shown in panels (a-c) of Fig. \ref{xz2} it becomes more than evident that the area of the stability islands is constantly being reduced with increasing energy. However for $E = -1$ we see that this tendency is reversed. Additional numerical calculation reveal that the regular orbits for $E = -1$ are somehow different with respect to those for $E < -1$. In all previous cases, the vast majority of non-escaping regular orbits corresponds to 1:1:0 resonant loop orbits, where the loop is parallel to the $(x,y)$ plane (see panel (a) of Fig. \ref{orbs0}). On the other hand, when $E = -1$ the initial conditions of the stability islands correspond to 1:1:0 tilted loop orbits. A characteristic example of such a regular 3D orbit is given in panel (b) of Fig. \ref{orbs0}. At this point, we would like to emphasize that similar tilted loop orbits have also been observed in barred galaxies \citep[see Fig. 21 in][]{JZ15}. In fact we found that inclined loop orbits seem to be the most persistent type of stable 3D orbits in barred galaxies.

The distribution of the escape time of the orbits on the $(x,z)$ plane for the values of the energy presented in Fig. \ref{xz2}(a-d) is shown in Fig. \ref{xzt2}(a-d), respectively. One may observe that the results are very similar to those presented earlier in Fig. \ref{xzt1}(a-d), where we found that orbits with initial conditions inside the basins of escape have the smallest escape rates, while on the other hand, the longest escape periods correspond to orbits with initial conditions either in the fractal regions of the plots, or near the boundaries of the stability islands of non-escaping regular motion.

For the numerical integration of the initial conditions of the three-dimensional orbits in each colour-coded grid on the $(x,z)$ plane, shown in Figs. \ref{xz1} and \ref{xz2}, we needed about 1 day of CPU time on a Quad-Core i7 2.4 GHz PC.

\begin{figure}
\begin{center}
\includegraphics[width=\hsize]{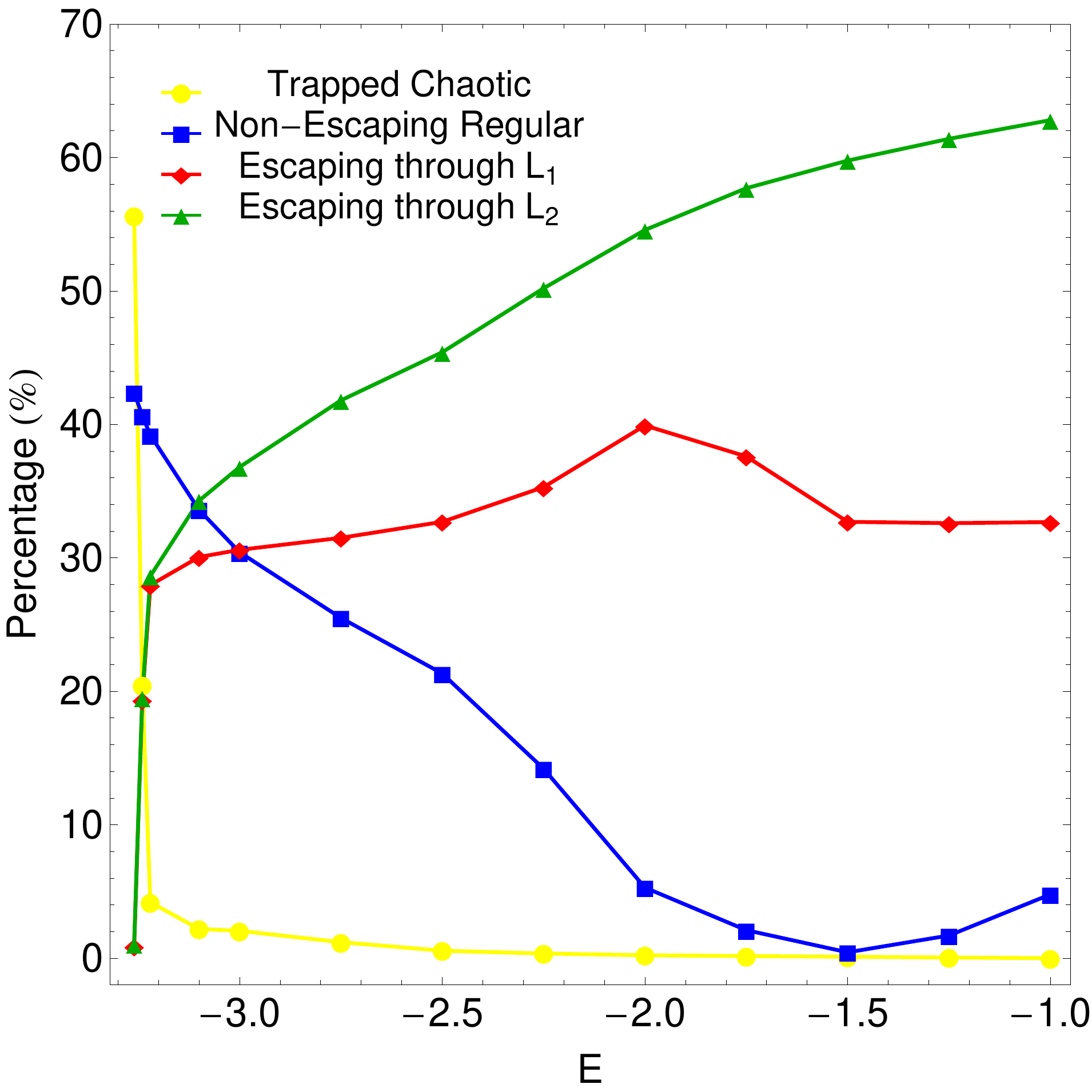}
\end{center}
\caption{Evolution of the percentages of all types of orbits on the $(x,z)$ plane, as a function of the total orbital energy $E$. The distribution shown applies to a particular plane of initial conditions. The asymmetry between the saddles $L_1$ and $L_2$ is caused by the choice of one branch of $p_{y0}$ (i.e. one orientation of the initial intersection of the plane $y = 0$). (For the interpretation of references to colour in this figure caption and the corresponding text, the reader is referred to the electronic version of the article.)}
\label{percs}
\end{figure}

\begin{figure}
\begin{center}
\includegraphics[width=\hsize]{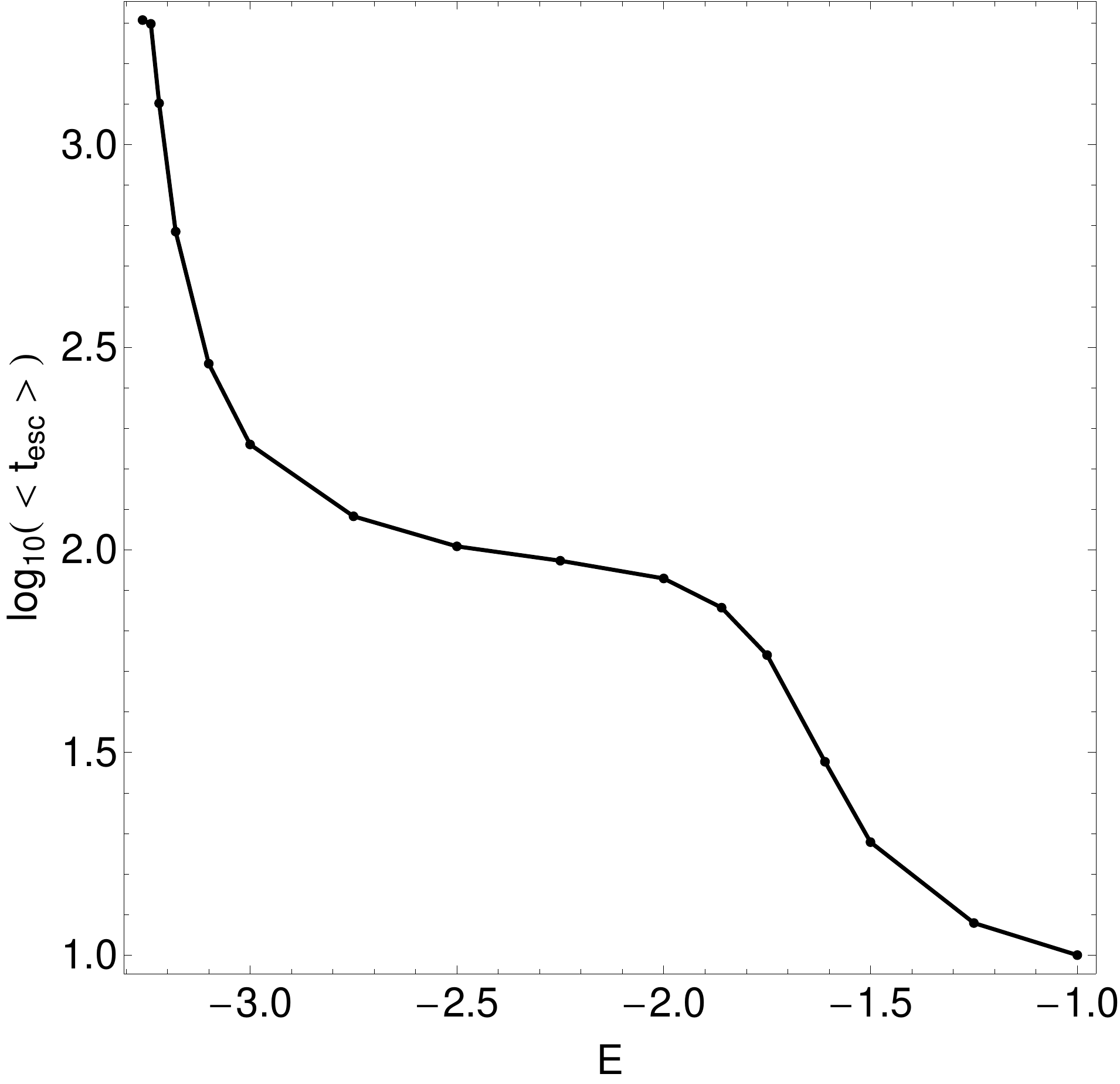}
\end{center}
\caption{Evolution of the logarithm of the average escape time of the orbits $(\log_{10}\left( < t_{\rm esc} > \right) )$, as a function of the total orbital energy $E$.}
\label{tesc}
\end{figure}

Undoubtedly, it would be very illuminating to monitor the evolution of the percentages of all types of orbits as a function of the total orbital energy $E$. The corresponding diagram is given in Fig. \ref{percs}. We would like to point out that for constructing this diagram we used data from additional colour-coded grids apart from those given in Fig. \ref{xz1} and Fig. \ref{xz2}. We observe that just above the energy of escape trapped chaotic motion dominates as the corresponding initial conditions of the orbits cover about more than half of the $(x,z)$ plane. However their percentage drops very quickly as the value of the energy increases. In particular, for $E > -2$ the rate of trapped chaotic orbits practically vanishes possessing extremely low values (lower than 0.01\%). Non-escaping regular orbits are also prominent at low energy levels above yet very close to the energy of escape. Our calculations suggest that the maximum percentage of regular orbits is about 42\% and it is observed for $E = -3.26$. With increasing energy the rate of non-escaping regular orbits decreases up to $E = -1.5$. At this energy level we observed the lowest rate of regular orbits which is about 1\%. We could say that at $E = -1.5$ we have the end of 1:1:0 loop orbits parallel to the $(x,y)$ plane. For higher values of the energy $(E > -1.5)$ the percentage of bounded ordered orbits starts to increase. However it should be noted that now the orientation of the 1:1:0 regular loop orbits has been changed (they become inclined to the $(x,y)$ plane). The evolution of the escaping orbits is completely different with respect to the bounded orbits (regular or chaotic). Looking at Fig. \ref{percs} one can see that for values of energy very close to the escape energy the rates of both exit channels are equal which of course means that both channels are equiprobable. On the other hand, for $E > -3.22$ the rates of escaping orbits start to diverge. Being more specific, the portion of escaping orbits through $L_2$ increases more rapidly with respect to the portion of escaping orbit through $L_1$. Furthermore, for $E > 1.5$ the rate of escapers through $L_1$ seems to saturates around 32\%, while the rate of escapers through $L_2$ continues to grow. At the highest energy level studied, that is $E = -1$, escaping orbit through exit channel 2 occupies about 63\% of the $(x,z)$ plane which is about twice the rate of escapers through exit channel 1. Therefore taking into account all the above-mentioned analysis we may conclude that at low energy levels, where the fractality of the $(x,z)$  plane is maximum, stars do not show any particular preference regarding the escape channels. On the contrary, at high enough energy levels, where basins of escape dominate, it seems that exit channel 2 (escape towards infinity) is twice more preferable with respect to exit channel 1 (escape towards the galactic center).

Of course, the relative fraction of orbits escaping through saddle $L_1$ and through saddle $L_2$ strongly depends on our particular choice of the collection of initial conditions. The inversion symmetry of the whole system mentioned before implies that to each orbit escaping through saddle $L_1$ there is another orbit with inverted initial conditions which escapes through saddle $L_2$. Therefore an integration over the whole 5-dimensional energy shell of possible initial conditions would give equal escape rates through both saddles. And also a random collection of initial conditions from the
interior part of the 5-dimensional energy shell would lead to equal escape rates. For grids of initial conditions on any particular lower dimensional surface these escape rates depend on how this particular surface intersects the basins of escape belonging to the two saddles.

Before closing this section, we would like to present in Fig. \ref{tesc} the evolution of the average value of the escape time $< t_{\rm esc} >$ of the orbits as a function of the total orbital energy $E$. We observe that very close to the critical energy of escape the average escape time of the orbits correspond to about 2000 time units. However, with increasing energy, the required time for escape smoothly reduced until $E = -2$, where $< t_{\rm esc} > \simeq 100$. For higher values of the energy the reduction of $t_{\rm esc}$ continues but with a different slope. For $E = -1$ the magnitude of the escape time of the orbits has been reduced over two orders with respect to that for $E = -3.26$. Indeed, for $E = -1$ the vast majority of the orbits need no more than 10 time units in order to escape.

In Paper II we provided an explanation for the observed behavior of the escape time of the orbits. We underlined that as we proceed to higher energy levels the two escape channels around the Lagrange points become more and more wide. Therefore, stars need less and less time to find one of the two symmetrical exits in the equipotential surface and eventually escape from the cluster. This simple geometrical feature explains in a very satisfactory way the energy-evolution of the escape time. In particular, it justifies why for low values of the energy orbits consume large time intervals wandering inside the limiting surface until they locate one of the two exits and escape either towards the galactic center or towards the infinity.

\section{The NHIM over the potential saddle}
\label{nhims}

When we want to analyse and understand the dynamics of a Hamiltonian system then in many cases it is easier to represent the system by its Poincar\'e map for fixed energy instead of its flow. For a 2-dof Hamiltonian system the Poincar\'e map acts on a 2-dimensional domain, therefore it can be presented by 2-dimensional graphics and gives an instructive overview of the dynamics for fixed energy. Even better, if we have the map for various values of the energy or of any other parameter, then we see the development scenario of the system in a graphical form. For Poincar\'e maps see the books on dynamical system theory \citet{J91} and \citet{LL93}, while pictorial explanations can be found in \citet{AS92}. In the present article we deal with a Hamiltonian 3-dof system and then the Poincar\'e map for fixed energy acts on a 4-dimensional domain and can no longer be presented by 2-dimensional graphics. Then the important question is: How to gain an understanding of the important
properties of this map? In addition, there is a more specific feature of these maps which is essential for all kinds of escape processes and which we like to transfer from the 2-dof to the 3-dof case. If the effective potential has an index-1 saddle, then the flow over this saddle is the essential step in the escape process and the escape is directed by some invariant subset sitting over this saddle and by its stable and unstable manifolds.

In 2-dof systems the invariant subset usually is an unstable periodic orbit which in the Poincar\'e map appears as hyperbolic fixed point. And the stable and unstable manifolds of this periodic orbit in the flow (which appears as a hyperbolic fixed point in the map) direct and channel the flow over the saddle. The important point is that the 2-dimensional stable and unstable manifolds of the orbit in the 3-dimensional flow for fixed energy (or the corresponding 1-dimensional stable and unstable manifolds of the fixed point in the 2-dimensional map) are of codimension 1 and therefore they divide the phase space into regions of different behaviour and are able to confine and channel the motion. The invariant subsets like the periodic orbit in the 3-dimensional flow or the fixed point in the domain of the map are of codimension 2.

When we go over to a 3-dof system and look for subsets of analogous properties, then the codimensions and the normal hyperbolicity are the essential properties which we must maintain. When we have some invariant subset with stable and unstable manifolds of codimension 1, then these manifolds again divide the phase space and are able to confine and channel the general motion. Accordingly, in a 4-dimensional map we need a 2-dimensional invariant subset sitting over the index-1 saddle of the effective potential and having stable and unstable manifolds of dimension 3. In addition, such invariant subsets should be persistent under general perturbations of the system in order to enable us to follow the scenario of the
dynamics under parameter changes, for example under a change of the energy. Subsets having all these properties are known to mathematics, they are the normally hyperbolic invariant manifolds (NHIMs) of codimension 2. They act as a kind of most important elements of the skeleton of the whole dynamics. Additional general information on NHIMs can be obtained in \citet{W94}. In the present section we will study the NHIM sitting over the Lagrange point $L_2$ of our system and in the following section we will study its implications for the tidal tails of the star cluster.

\subsection{The NHIM and the Lyapunov orbits over an index-1 saddle }
\label{ns1}

First let us consider the orbits which stay forever over an index-1 saddle like the Lagrange point $L_2$. We will describe the important properties in a descriptive form without going into detailed calculations. The considerations of this subsection are quite general and apply for any index-1 saddles of the effective potential in a rotating frame, for example they coincide with the corresponding properties of the NHIM over the Lagrange points for a barred galaxy as described in Paper III. In particular, the equations given in the beginning of section 4 and in the subsections 4.1 and 4.2 of Paper III can be taken over for the present case immediately. Therefore we do not repeat these equations here and we only give a detailed description of important properties. In the present article we shall use the same notation as in Paper III, as far as possible.

When the total orbital energy (numerical value of the Hamiltonian in the rotating frame of reference) is above but close to the saddle energy, then the dynamics over the saddle is well approximated by a quadratic expansion of the Hamiltonian in the phase space coordinates. In this approximation the equations of motion are linear and accordingly there are three normal modes of motion in the neighbourhood of the saddle. Since we have an index-1 saddle, one of these normal modes is unstable the other ones are stable. The first stable mode is a vertical oscillation. When all available energy goes into this particular motion then we have the vertical Lyapunov orbit, called $\Gamma_v$ in the following. The second stable mode is motion along a horizontal ellipse. When all available energy goes into this particular motion then we have the horizontal
Lyapunov orbit, called $\Gamma_h$ in the following. The third mode is unstable horizontal motion which falls down exponentially along the unstable direction of the effective potential. The general orbit stays in the neighbourhood of the saddle point for infinite time in the past and in the future if and only if it is a superposition of the two stable modes and does not contain any contribution of the unstable motion of the third mode.

For fixed total orbital energy there are two free parameters for the general orbit which stays over the saddle. First, we can distribute the available energy between the two stable modes. Second, we can choose any phase shift between these two stable
modes. Accordingly, there is a 2-dimensional continuum of orbits which remain in the neighbourhood of the saddle for ever. This continuum forms a 3-dimensional surface in the 5-dimensional flow for fixed energy or a 2-dimensional surface in the 4-dimensional domain of the map. These surfaces are of codimension 2 and they are unstable (hyperbolic) in the directions normal to the surface, i.e. these surfaces are the NHIMs we have been looking for. For energy sufficiently close to the saddle energy the topology of the NHIM surface in the flow is the one of the 3-dimensional sphere $S^3$. The NHIM of the map is represented easiest as a curved disc. However, by contracting the boundary of the disc to a single point we can also
represent it as a 2-dimensional sphere $S^2$ whenever this representation is more convenient.

What happens under perturbations, for example when the energy increases and nonlinearities in the dynamics become important? Here one extremely important property of NHIMs becomes essential, namely their persistence under perturbations. With nonlinearities also the dynamics restricted to the invariant NHIM surface can develop instabilities. But for small perturbations the tangential instabilities are smaller than the normal instability and then the NHIM surface survives as invariant surface. It may be shifted a little or deformed smoothly but it conserves its topology and conserves its NHIM properties. In particular, this implies also the persistence of its stable and unstable manifolds. Only for large perturbations the normal hyperbolicity can be lost and then the NHIM may change qualitatively, may change its topology or may decay and be lost. The reader can find more information regarding the persistence properties in \citet{BB13,E13,F71,W94}, and for the bifurcations of NHIMs under perturbations in \citet{AB12,MS14,MCE13,TTK11,TTK15,TTT15}. Next we will investigate the perturbation scenario of the NHIM in our system under an increase of the energy.

\subsection{The development scenario of the Lyapunov orbits}
\label{ns2}

The most prominent periodic orbits inside of the NHIM are the Lyapunov orbits and some further periodic orbits related to the Lyapunov orbits. In the present subsection we shall study the development scenario of these periodic orbits under an increase of the energy as preparation of the study of the development scenario of the NHIM.

\begin{figure*}
\centering
\resizebox{\hsize}{!}{\includegraphics{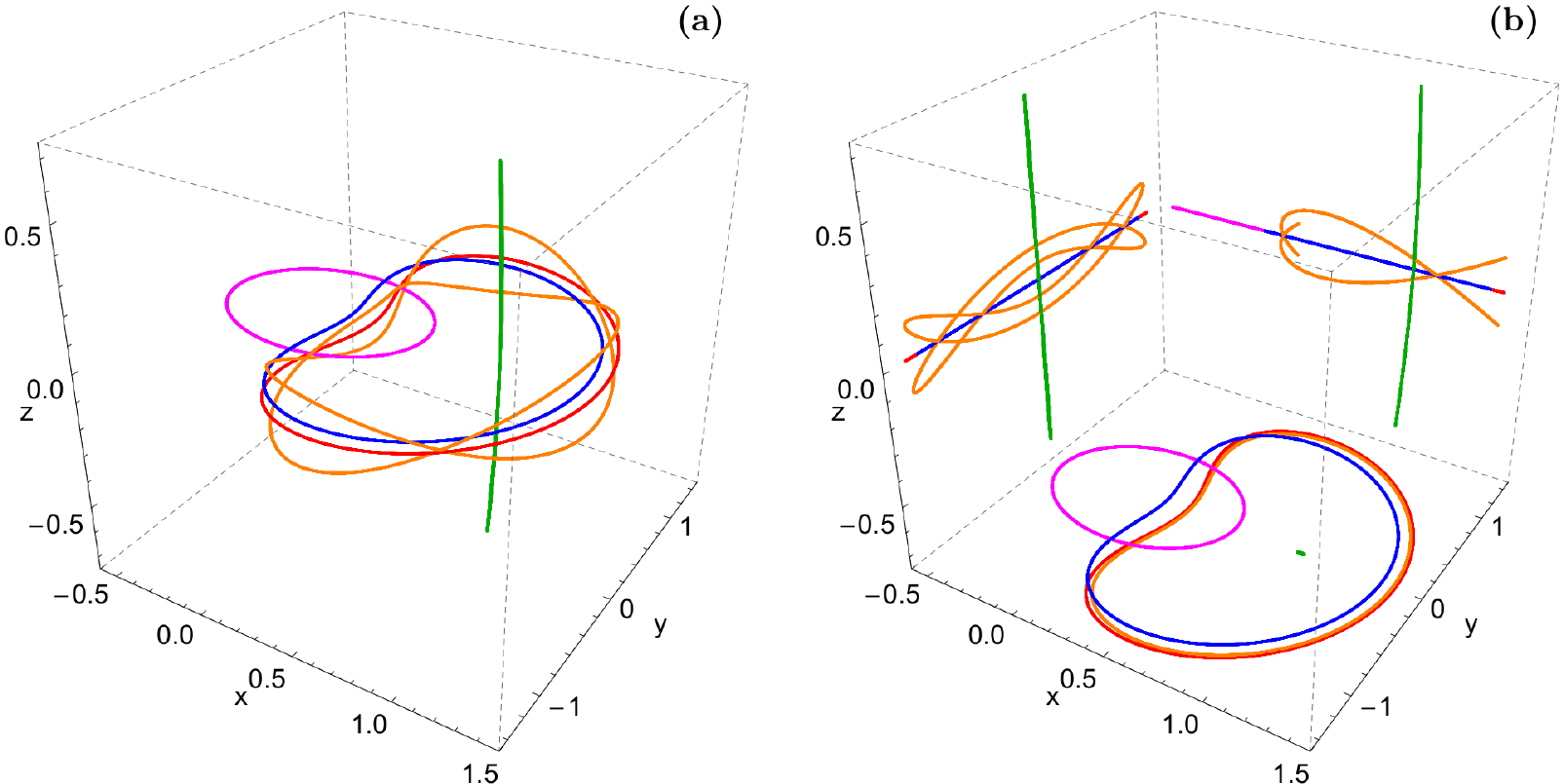}}
\caption{(a-left): A collection of the most important periodic orbits in the configuration $(x,y,z)$ space, for the energy $E = -1.160$. (b-right): The projections of the periodic orbits into the primary planes $(x,y)$, $(x,z)$ and $(y,z)$. The colour code is as follows: orbit $\Gamma_v$ (green), orbit $\Gamma_r$ (magenta), orbit $\Gamma_s$ (blue), orbit $\Gamma_h$ (red), orbits $\Gamma_{t1}$ and $\Gamma_{t2}$ (orange). (For the interpretation of references to colour in this figure caption and the corresponding text, the reader is referred to the electronic version of the article.)}
\label{orbs}
\end{figure*}

\begin{figure}
\includegraphics[width=\hsize]{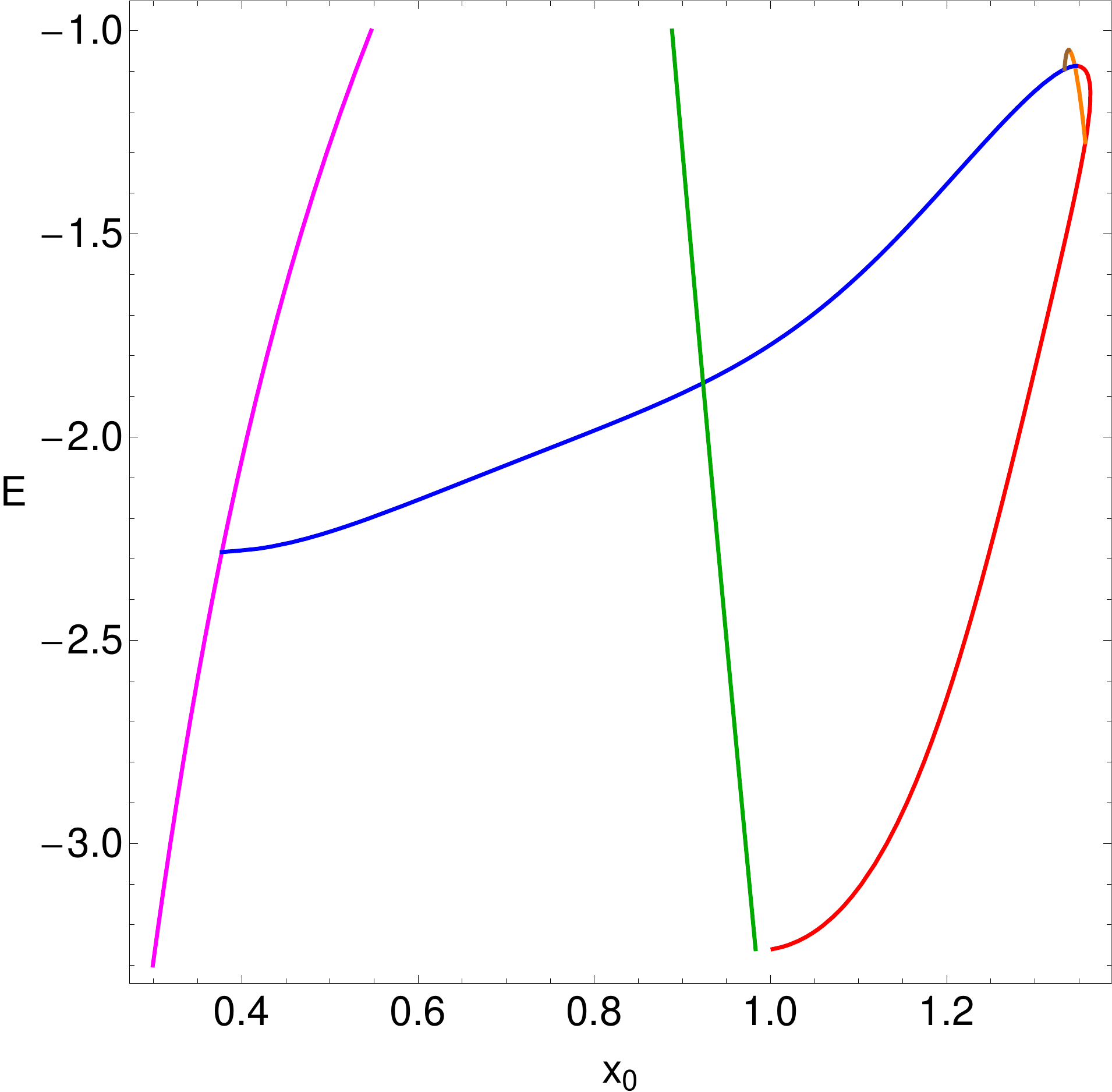}
\caption{Evolution of the $x_0$ initial conditions of the periodic orbits as a function of the orbital energy $E$. The colour code is the same as in Fig. \ref{orbs}, while orbits $\Gamma_{u1}$ and $\Gamma_{u2}$ are shown in brown. (For the interpretation of references to colour in this figure caption and the corresponding text, the reader is referred to the electronic version of the article.)}
\label{fpo}
\end{figure}

In our system the vertical Lyapunov orbit $\Gamma_v$ follows an extremely simple scenario. The orbit $\Gamma_v$ is born at the saddle energy $E_L$ and is stable in the directions tangential to the NHIM and unstable in the directions normal to the NHIM. The qualitative stability properties do not change under an increase of $E$, only the numerical values of the eigenvalues change slowly. Accordingly, the orbit $\Gamma_v$ does not suffer any bifurcations along the whole scenario. When we use the words tangential and normal here and in the following then it always refers to the NHIM surface.

Also the horizontal Lyapunov orbit $\Gamma_h$ is born at the saddle energy as tangentially stable and normally unstable. Near the energy $E = -1.275 $ it suffers a pitchfork bifurcation where it splits off a pair of tilted loop orbits, called $\Gamma_{t1}$ and $\Gamma_{t2}$ in the following. The orbit $\Gamma_v$ itself becomes tangentially unstable in this bifurcation. However, its tangential instability remains small compared to its normal instability. Accordingly, orbit $\Gamma_h$ remains part of the NHIM. Near $E = -1.0872$ the Lyapunov orbit $\Gamma_h$ collides with another periodic orbit called $\Gamma_s$ in the following which is normally elliptic and the two colliding orbits disappear in a saddle-centre bifurcation in normal direction.

The orbit $\Gamma_s$ itself is split off near $E = -2.283$ in a pitchfork bifurcation from another periodic orbit $\Gamma_r$ described below. The orbit $\Gamma_s$ is born stable in all directions and under increasing energy it changes its stability properties several times while it runs through several bifurcations where only two of them are of interest for us. Near $E = -1.095$ in a pitchfork bifurcation the orbit $\Gamma_s$ splits off two further tilted loop orbits which we call orbits $\Gamma_{u1}$ and $\Gamma_{u2}$ in the following. And finally, near $E = -1.0872$ it disappears, as mentioned above.

The periodic orbit $\Gamma_r$ mentioned above comes out of the potential hole around the origin and it encircles the origin in a shape which is symmetric under the reflection $x \to - x$. For small values of the energy it is stable in all directions. Near $E = -2.283$ it suffers a pitchfork bifurcation where it becomes unstable in one plane and splits off two
new periodic orbits where one of them is the orbit $\Gamma_s$ mentioned above. The orbit $\Gamma_s$ is no longer invariant under the reflection $x \to - x$, however under this reflection the orbit $\Gamma_s$ is mapped into the second periodic orbit split off from the orbit $\Gamma_r$ in the pitchfork bifurcation. Under increasing energy orbit $\Gamma_s$ moves in direction of increasing values of $x$, i.e. it moves toward the saddle point $L_2$. The symmetry counterpart moves toward point $L_1$.

The tilted loop orbits $\Gamma_{t1}$ and $\Gamma_{t2}$ are born normally unstable and tangentially stable. They are part of the NHIM. Near $E = -1.0481$ the orbits $\Gamma_{t1}$ and $\Gamma_{t2}$ collide with the other tilted loop orbits $\Gamma_{u1}$ and $\Gamma_{u2}$ respectively and all these tilted loop orbits disappear in two symmetry related saddle-centre bifurcations.

\begin{figure*}
\centering
\resizebox{\hsize}{!}{\includegraphics{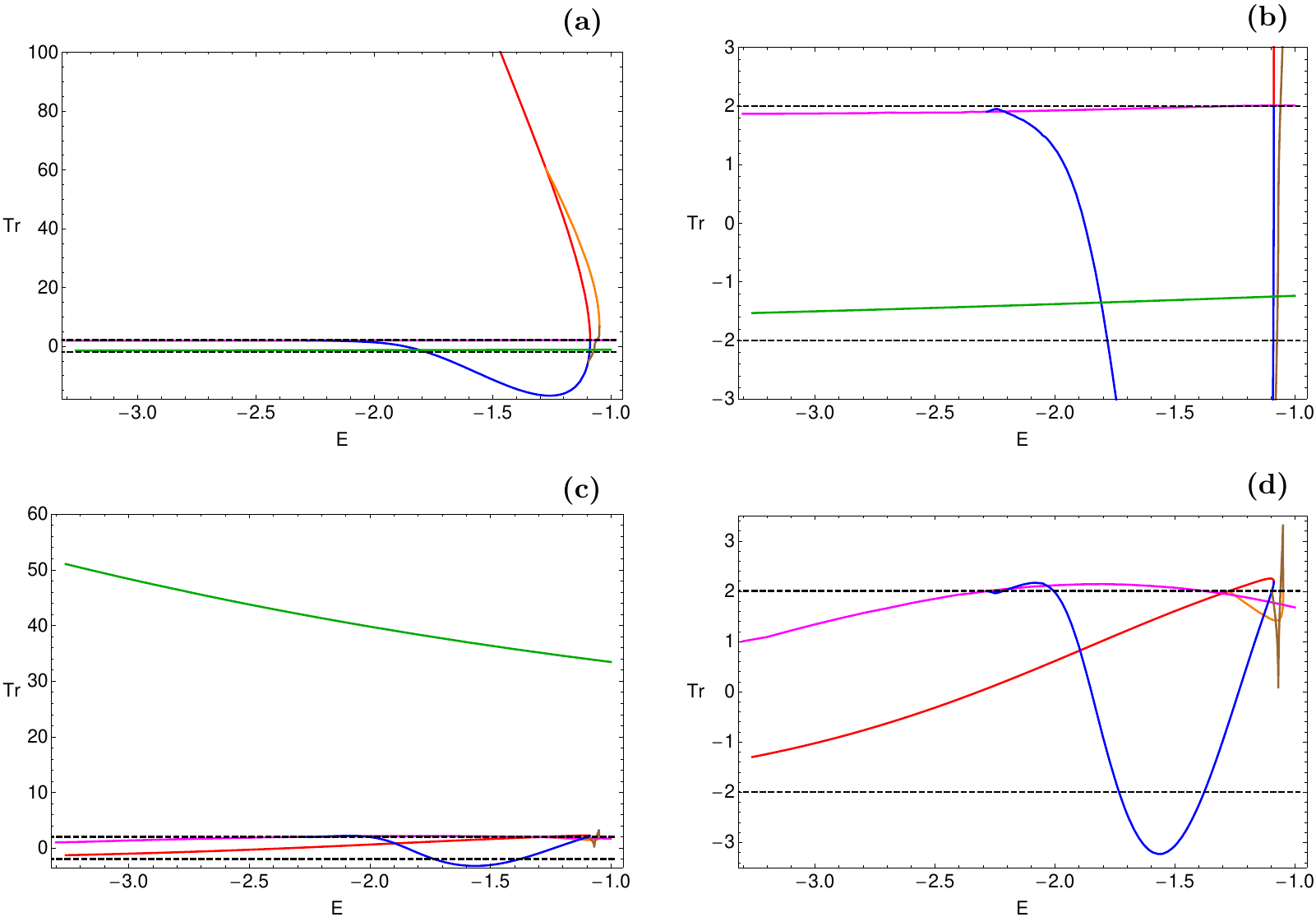}}
\caption{Evolution of (a upper left): the normal and (c lower left): the tangential traces of the monodromy matrix of the families of periodic orbits as a function of the total orbital energy $E$. Magnifications of panels (a) and (c) are shown in panels (b) and (d), respectively. The colour code is the same as in Fig. \ref{fpo}. The horizontal black dashed lines correspond to the critical values $\rm Tr_{\rm crit} = \pm 2$, which distinguish between stable and unstable motion. (For the interpretation of references to colour in this figure caption and the corresponding text, the reader is referred to the electronic version of the article.)}
\label{trs}
\end{figure*}

In the following plots we present this scenario graphically. Fig. \ref{orbs}(a-b) shows the mentioned periodic orbits in the position space for the energy level $E = -1.16$. Note that for this energy the orbits $\Gamma_{u1}$ and $\Gamma_{u2}$ do not yet exist. Panel (a) of Fig. \ref{orbs} is a perspective view in the 3-dimensional $(x,y,z)$ position space, while panel (b)
gives the 3 projections into the 3 different coordinate planes. Green colour represents the orbit $\Gamma_v$, magenta colour represents the orbit $\Gamma_r$, blue colour represents the orbit $\Gamma_s$, red colour represents the orbit $\Gamma_h$, while the two tilted loop orbits, $\Gamma_{t1}$ and $\Gamma_{t2}$, are both plotted in orange colour. Note how the red and the blue curve already become similar, so we can already anticipate the collision between these two orbits and their destruction in the saddle-centre bifurcation under a small further increase of the energy. Also note how close the tilted loop orbits come to the orbit $\Gamma_h$ in the projection into the $(x,y)$ plane. This illustrates how the tilted loop orbits are split off from $\Gamma_h$ in a pitchfork in $z$-direction. We also see very well how the orbit $\Gamma_{t1}$ is obtained from the orbit $\Gamma_{t2}$ by a reflection in the plane $z = 0$.

In Fig. \ref{fpo} we show as function of the energy the $x$ coordinate of all these orbits at the moment of intersection of the plane $y = 0$ in negative orientation. The use of colours coincides with the one of Fig. \ref{orbs}, while the additional orbits $\Gamma_{u1}$ and $\Gamma_{u2}$ are represented by the brown curve. Observe that a pair of tilted loop orbits gives a single curve only because of the above mentioned symmetry under the reflection $z \to - z$.

The stability properties of the periodic orbits as function of the energy are presented in Fig. \ref{trs}(a-d). For all orbits, with the exception of the orbits $\Gamma_{u1}$ and $\Gamma_{u2}$, the 4-dimensional monodromy matrices have a natural decomposition into two 2-dimensional blocks and then we plot the traces of these $2 \times 2$ blocks. For such orbits which belong to the NHIM we plot in panels (a) and (b) the trace belonging to the plane normal to the NHIM, and in panels (c) and (d) the trace belonging to the plane tangential to the NHIM. For the other orbits which do not belong to the NHIM we distribute the 2 traces to the two sets of plots according to their natural connection to the normal and tangential planes in the moment of bifurcations with other orbits belonging to the NHIM. The orbits $\Gamma_{u1}$ and $\Gamma_{u2}$ run through a complex spiralling case in the energy interval $E \in [-1.0687, -1.0491]$. Then the monodromy matrix does not have a splitting into 2 real $2 \times 2$ blocks, there is only a complex trace belonging to a splitting into two complex conjugate $2 \times 2$ blocks. In this case we plot into all panels of Fig. \ref{trs} the absolute value of this complex trace. Panel (b) is a magnification of panel (a) and panel (d) is a magnification of panel (c) in order to show better the important part of the small trace values. The important values +2 and -2 are marked by horizontal black dashed lines since these are the values of the traces where the bifurcations occur. The relation between the various orbits and the colours is as in Fig. \ref{fpo}.

\begin{figure*}
\centering
\resizebox{\hsize}{!}{\includegraphics{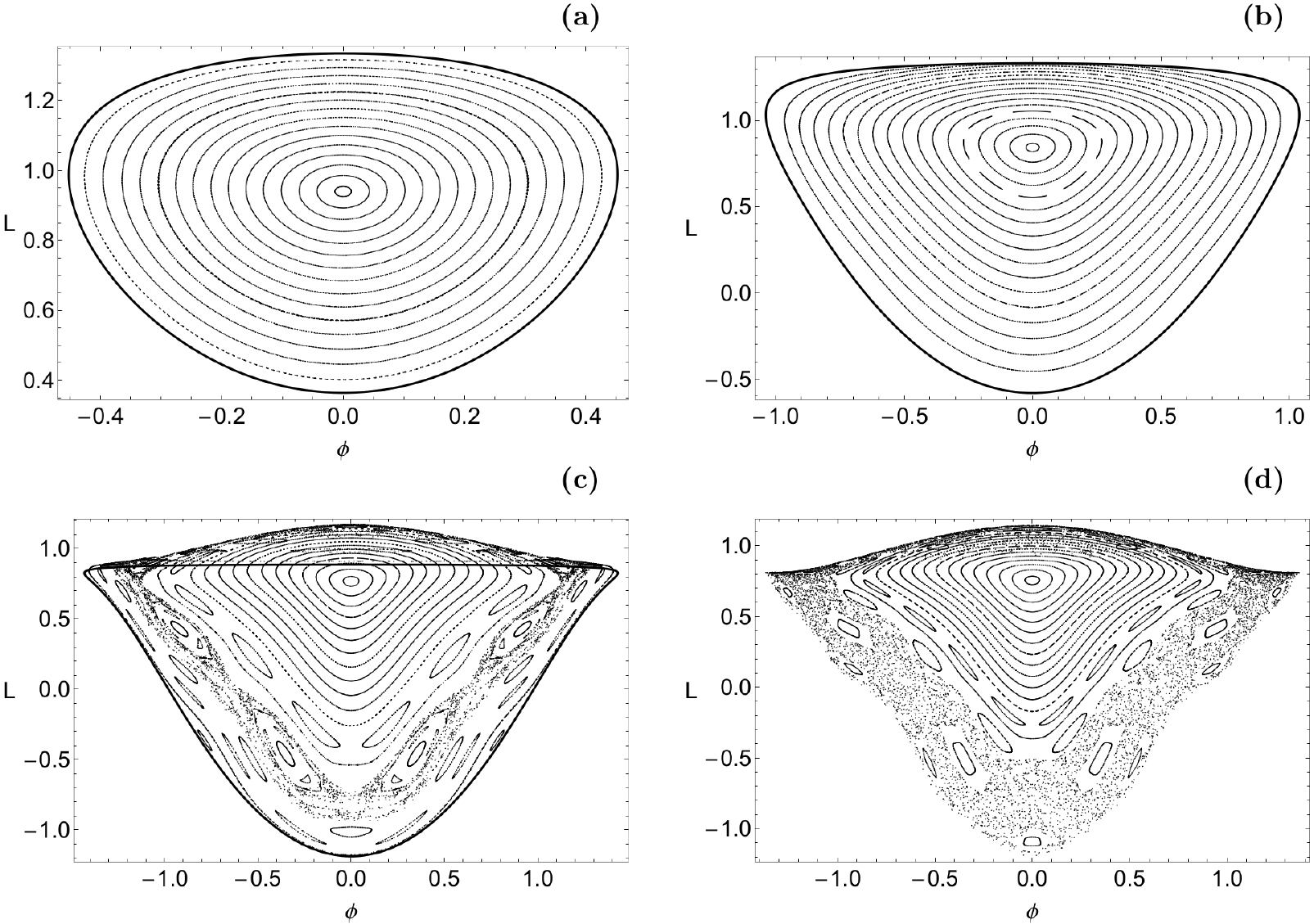}}
\caption{Projections of the NHIM surfaces into the $(\phi, L)$ plane. The outermost solid closed curve corresponds to the horizontal Lyapunov orbit. (a-upper left): $E = -3$; (b-upper right): $E = -2$; (c-lower left): $E = -1.15$; (d-lower right): $E = -1$.}
\label{maps}
\end{figure*}

\begin{figure*}
\centering
\resizebox{\hsize}{!}{\includegraphics{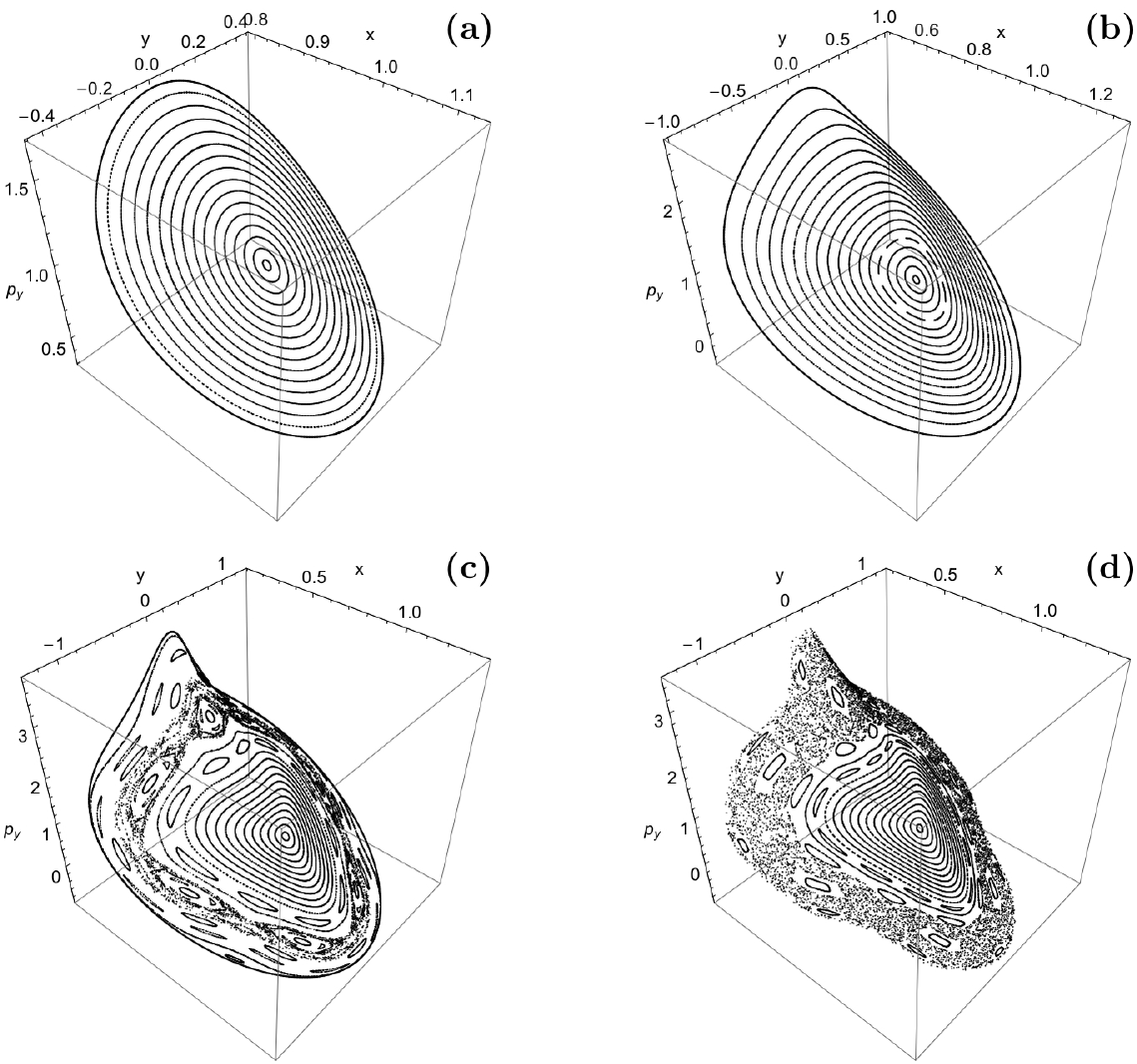}}
\caption{Projections of the NHIM surfaces into the $(x,y,p_y)$ phase space. The outermost solid closed curve corresponds to the horizontal Lyapunov orbit. (a-upper left): $E = -3$; (b-upper right): $E = -2$; (c-lower left): $E = -1.15$; (d-lower right): $E = -1$.}
\label{maps3d}
\end{figure*}

Note that one trace of the orbit $\Gamma_r$ is always rather close to +2. This orbit runs close to the origin basically in the almost rotationally symmetric potential of the interior of the star cluster. This approximate rotational symmetry implies approximate neutral stability under a rotation of the plane of this orbit and correspondingly it implies an approximate value 2 of the trace. Also note that the stability properties of a pair of tilted loop orbits coincide because of their interchange symmetry under the reflection $z \to - z$.

\subsection{The numerical restricted map on the NHIM}
\label{ns3}

The NHIM in the map is a 2-dimensional invariant subset of the domain of this map. This implies that the iterated images and preimages of any initial point from this subset belong to the same subset. Therefore the restriction of the Poincar\'e map to this NHIM surface makes sense and it is a 2-dimensional Poincar\'e map which can be represented by 2-dimensional graphics. For previous examples of the construction and use of this restricted map see \citet{GDJ14,GJ15} and Paper III. Since NHIMs are persistent under perturbations also the existence of this map is persistent under perturbations and is an interesting tool to study the perturbation scenario in 3-dof systems by 2-dimensional graphics. It represents graphically the development scenario of the NHIM and it embeds the development scenario of the Lyapunov orbits studied in the previous subsection into the development scenario of the whole NHIM. For the idea how such a restricted map is constructed numerically see \citet{GDJ14}.

For our numerical examples of Poincar\'e maps we always use the intersection condition $z = 0$ with positive orientation. As usual for the graphical presentation of Poincar\'e maps we choose a moderate number of initial points in the domain of the map and plot a large number of iterates of these initial points. The NHIM surface is a curved surface embedded into a 4-dimensional space. So to plot the map on the surface we must use some kind of projection. In analogy to Paper III it was instructive to use on one hand the projection on the $(\phi, L)$ plane where $\phi = \arctan(y/x)$ and $L = x p_y - y p_x$ and on the other hand perspective views of projections into the 3-dimensional $(x, y, p_y)$ space. This perspective view gives a suggestion of the curvature of the NHIM surface in the higher dimensional embedding space.

For energy close to the saddle energy the dynamics comes close to a linear dynamics and therefore the restricted map is similar to the Poincar\'e map of a 2-dof anisotropic harmonic oscillator. As a numerical example of this case see panel (a) of Fig. \ref{maps} which is constructed for $E = -3$ and uses the projection into the $(\phi, L)$ plane. The corresponding perspective plot in the $(x, y, p_y)$ space is shown in panel (a) of Fig. \ref{maps3d}. The central fixed point on the NHIM represents the Lyapunov orbit $\Gamma_v$ at the moment of its intersection with the plane $z = 0$ and the boundary of the domain is the Lyapunov orbit $\Gamma_h$ which lies completely in the plane $z = 0$. If we want to represent also the orbit $\Gamma_h$ by a fixed point of this map then we can contract the boundary to a single point. Thereby the domain of the map becomes a closed surface with the topology of $S^2$. The rest of the domain of the map is filled by invariant lines which encircle the central fixed point. They represent the quasi-periodic superposition of vertical motion along the Lyapunov orbit $\Gamma_v$ and horizontal motion along the Lyapunov orbit $\Gamma_h$. From the inside to the outside the energy in the vertical motion decreases and the energy in the horizontal motion increases. For the central fixed point all energy is in the vertical motion and for the boundary all energy is in the horizontal motion.

When we increase the energy to the value $E = -2$ little changes as panels (b) of the Figs. \ref{maps} and \ref{maps3d} indicate. This is consistent with the absence of bifurcations of the Lyapunov orbits in the energy interval $[-3,-2]$. Panels (c) of the same figures show the restricted map for $E = -1.15$. Now we are in the energy range of intermediate perturbation. In particular near the boundary many secondary structures have become visible. In particular note the new periodic points near $\phi = \pm 0.8$, $L = -0.35$. These two new structures belong to the two tilted loop orbits $\Gamma_{t1}$ and $\Gamma_{t2}$ which are split off from the Lyapunov orbit $\Gamma_h$, i.e. from the boundary, at $E = -1.275$ in a pitchfork bifurcation. Finally, in the panels (d) of Figs. \ref{maps} and \ref{maps3d} the energy has advanced to the value $E = -1$. Now the Lyapunov orbit $\Gamma_h$ and the tilted loop orbits have completely disappeared and the NHIM surface has lost its outer boundary.

On the other hand, the inner part of the surface around the fixed point belonging to the Lyapunov orbit $\Gamma_v$ remains unaffected during the whole scenario. This is completely consistent with the property of the orbit $\Gamma_h$ not to suffer any bifurcations and to remain normally hyperbolic. This remains so up to high positive energies, and therefore also a part of the NHIM around the orbit $\Gamma_h$ remains for these high energies.

After having seen the development scenario of the Lyapunov orbits and of the NHIM in our present system, the following questions naturally arise: Is this scenario well known? Does it fit into a more general scheme? How does it compare to the scenarios in similar systems?

\begin{figure*}
\centering
\resizebox{\hsize}{!}{\includegraphics{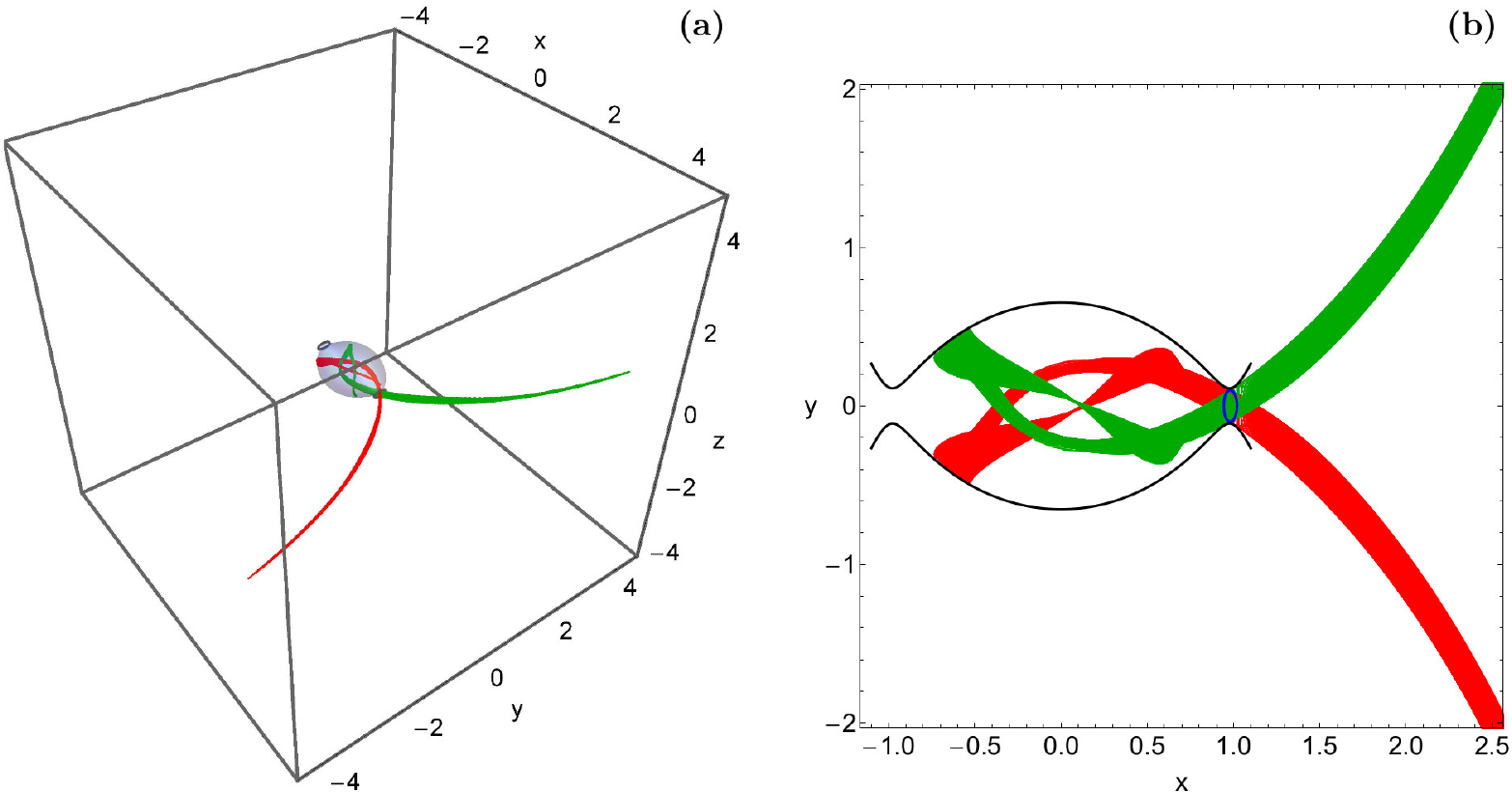}}
\caption{(a-left): The stable manifold $W^s(S_E)$ (green) and the unstable manifold $W^u(S_E)$ (red), when $E = -3.24$. The corresponding energetically allowed region of the position space in the inner part of the potential is shown in transparent gray colour. (b-right): The projection of the stable and the unstable manifold on the configuration $(x,y)$ space. The corresponding horizontal Lyapunov orbit $\Gamma_h$ is shown in blue. (For the interpretation of references to colour in this figure caption and the corresponding text, the reader is referred to the electronic version of the article.)}
\label{mans}
\end{figure*}

Unfortunately, the investigation of the general development scenarios of NHIMs is a problem little explored up to now. The reader can find additional information on this interesting problem in the series of papers \citet{AB12,MS14,MCE13,LST06,TTK11,TTK15,TTT15}. Therefore, it is impossible to say, at this moment, whether or not our present scenario fits into any general scheme. More specifically we can compare our present scenario to the scenarios of other systems in a rotating frame of reference and having Lagrange points which are index-1 saddles of the effective potential. Here 2 examples come to mind: First, the restricted three-body system as treated in \citet{JM99}, and second, the escape from a barred galaxy as it was described in Paper III. In \citet{JM99} the authors constructed the dynamics restricted to the centre manifold of the saddle and studied the Poincar\'e map of this restricted dynamics. Also they used the intersection condition $z = 0$ for their map. The restriction of the dynamics to the centre manifold and the restriction of the dynamics to a NHIM are the same basic idea. Thus, we can directly compare their restricted map to our restricted map. For instance, see Figs. 3 and 6 in \citet{JM99}. Also in this system the vertical Lyapunov orbit does not suffer any bifurcation and the horizontal Lyapunov orbit splits off a pair of tilted loop orbits. In this sense, the system of Jorba and Masdemont and the star cluster model run through the same basic scenario.

In contrast, the scenario of the barred galaxy is qualitatively very different. In the barred galaxy model the vertical Lyapunov orbit runs through various bifurcations and splits off two different pairs of tilted loop orbits. In this system the horizontal Lyapunov orbit does not split off tilted loop orbits. However, it splits off further horizontal orbits in a pitchfork. Certainly the star cluster and the barred galaxy belong to clearly distinct scenarios. The best way to make progress in the extremely difficult but also important problem of the development scenarios of NHIMs might be to study many examples, to recognize repeating patterns and to find phenomenological classifications. We hope that our present example gives a useful contribution to this program.

\subsection{The stable and unstable manifolds of the NHIM}
\label{ns4}

As mentioned above, the NHIMs have stable and unstable manifolds of codimension 1 and these invariant manifolds of the NHIMs are responsible for the importance of the NHIMs with respect to the global dynamics of the system. If $S_E$ denotes the NHIM surface for the energy value $E$ then we use the symbols $W^s(S_E)$ and $W^u(S_E)$ for its stable and unstable manifolds, respectively. $W^s(S_E)$ and $W^u(S_E)$ have each two branches, one going outwards from the saddle and one going inwards from the saddle respectively. $W^s(S_E)$ contains such orbits which converge towards $S_E$ in the future and $W^u(S_E)$ contains such orbits which converge towards $S_E$ in the past. This property gives already the idea how one can trace out $W^s(S_E)$ and $W^u(S_E)$ numerically. One simply takes a lot of initial conditions close to $S_E$ and let the orbits run. In the past
direction they approach automatically $W^s(S_E)$ and in the future they approach $W^u(S_E)$.

This is how panel (a) of Fig. \ref{mans} has been constructed for the energy $E = -3.24$. $W^s(S_E)$ is plotted in green colour, while $W^u(S_E)$ is plotted red. Of course, the manifolds live in the 5-dimensional energy shell of the phase space. The figure shows a perspective view of their projection into the 3-dimensional position space. The energetically allowed region of the position space in the inner part of the potential is marked by gray colour. The figure shows the local part of the manifolds only, i.e. the orbits have been integrated for a moderate time only. Globally, i.e. for large times, these manifolds grow complicated folds and tendrils. In panel (b) of Fig. \ref{mans} the projection of the stable and the unstable manifolds on the $(x,y)$ plane are given. It is seen that the horizontal Lyapunov periodic orbit $\Gamma_h$ exist in the intersection of the two manifolds. This periodic orbit is included into the same figure as the blue solid curve.

Fig. \ref{mans} shows the manifolds of the NHIM over the saddle point $L_2$ only. Because of the inversion symmetry of the system mentioned before, there is an equal NHIM over the saddle $L_1$. And one of these two NHIMs is obtained from the other one by a rotation around the $z$-axis by an angle $\pi$. Of course, the stable and unstable manifolds of one of these NHIMs
are also obtained from the stable and unstable manifolds of the other one by this symmetry operation. Globally the stable manifolds of one NHIM intersect the unstable manifold of the same NHIM and also the unstable manifold of the other NHIM. Such intersections represent homoclinic and heteroclinic orbits which converge in the past as well as in the future to one of the NHIMs. Close to such intersections exists also an infinity of periodic orbits which oscillate between the saddle points. General orbits move irregularly between these tangles.

For an understanding of the escape process the following picture of the behaviour of typical orbits starting in the inner part of the potential is helpful. The orbit moves in the inner potential hole and at some time it finds itself close to the local segment of the stable manifold of some saddle NHIM. Then it moves along this stable manifold to the neighbourhood of the NHIM and thereby also close to the saddle. When the orbit does not start exactly on the stable manifold then it stays in the neighbourhood of the NHIM for a finite time only and leaves it again close to its unstable manifold. It depends on fine details of the initial conditions whether it leaves along the inner or the outer branch, i.e. whether it returns to the inner potential hole or escapes to the outer region. The boundary between escape and return is the stable manifold of the NHIM itself. By the mechanism just described the NHIM is responsible for escape properties and for structures formed by the escaping orbits. More on this point will be presented in the next section.

In the previous subsection we studied the dynamics on the NHIM surface itself. Now one can ask whether this restricted dynamics has any implications for the dynamics outside of the NHIM. Here the stable and the unstable manifolds of the NHIM are responsible for an affirmative answer. We can imagine $W^s(S_E)$ and $W^u(S_E)$ as a kind of cylinders over the NHIM which have an internal foliation according to the structures on the NHIM itself. And this foliation carries the structures found on the NHIM to far away regions reached by $W^s(S_E)$ and $W^s(S_E)$. In this sense, the scenario found in the restricted dynamics on the NHIMs is related to the dynamics in far away regions influenced by $W^s(S_E)$ and $W^u(S_E)$.

\section{Formation of tidal tails}
\label{tdt}

This section is devoted to the fate of escaping stars. In particular, we will discuss what happens to stars when they pass either through $L_1$ or $L_2$ on their way to escape from the star cluster. It is well known that the majority of the stars of a cluster escape from it, through the Lagrange points, with relatively low velocities\footnote{Most stars escape as a result of two-body encounters. These escapers usually have an energy only slightly above the energy of escape \citep{H70}. Therefore, such escapers pass through one of the two Lagrange points at slow speed.}. Escaping stars usually form extended tidal tails (also known as tidal arms) \citep[e.g.,][]{CMM05,dMCM05,JBPE09,KMH08,KKBH10}. Stars in tidal tails move under the gravitational influence of the axially symmetric potential of the parent galaxy due to the fact that the attraction of the star cluster is practically negligible outside the tidal radius. Thus, due to the existence of Coriolis and centrifugal forces escaping stars display a simple epicyclic motion along the two tidal tails.

If there are stars in the interior region of the cluster with an energy high above the threshold energy $E(L_2)$ then such
stars will leave the interior region fast and the interior region will have lost such stars long time ago. Let us now consider stars with an energy below the threshold but close to it and moving in the inner part of the cluster. Such stars
have occasional interactions among themselves and with other objects and thereby their energy can be changed slightly and
it may come a little above the threshold. Then such stars are exactly the ones for which the structure of the NHIMs over the Lagrange points $L_1$ and $L_2$ and their stable and unstable manifolds become highly relevant.

First, these stars can come close to the saddle points along the stable manifolds of the NHIMs, and then they have two
possibilities for their further motion. First, from the neighbourhood of the saddle they can return to the inner region of the cluster along the inner branches of the unstable manifolds of the NHIMs. Second, they can leave to the outside of the cluster along the outer branches of the unstable manifolds of the NHIMs. Which one of these two possibilities is realised depends on which side of the local segment of its stable manifold the orbit approaches the NHIM. Stars which return to the inside will come back to the neighbourhood of a saddle point later and can eventually escape.

When an orbit starts in the neighbourhood of the saddle point $L_2$ then in forward direction (i.e. in the future) it converges automatically against $W^u(S_E)$. This observation provides the idea for a numerical construction of $W^u(S_E)$. We randomly select 1000 initial conditions close to $L_2$ all with the same energy $E = -3$ and we let the orbits run for a finite time. The easiest is to take these initial conditions from the corresponding maps presented in Figs. \ref{maps} and \ref{maps3d} (see panels (a)). Note that all these orbits start with $z = 0$. Approximately half of these orbits leave to the outside immediately along the outer branch of $W^u(S_E)$. The other half goes to the inside along the inner branch of $W^u(S_E)$. Later after several revolutions in the inside this half of the orbits escapes partly over $L_2$ and partly over $L_1$. In total all these orbits together trace out the outer unstable manifolds of both the NHIMs over $L_1$ and $L_2$.

\begin{figure}
\includegraphics[width=\hsize]{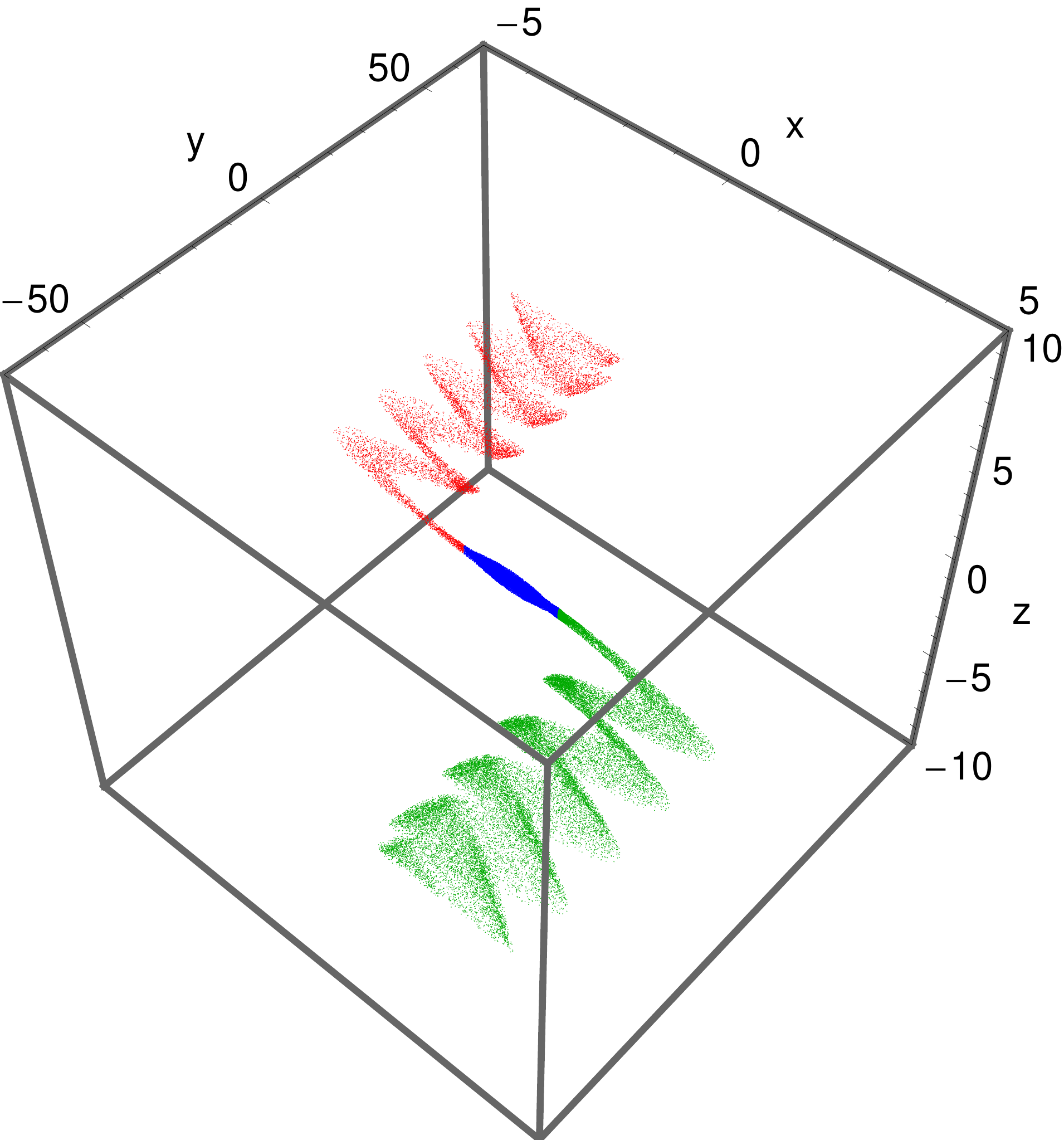}
\caption{The projection of the local segments of $W^u(S_E)$ into the $(x,y,z)$ position space, when $E = -3$. The manifolds from the NHIM over the saddle point $L_1$ are shown in red colour, the ones from the symmetrically placed NHIM over $L_2$ are shown in green colour, while stars inside the interior region are shown in blue. (For the interpretation of references to colour in this figure caption and the corresponding text, the reader is referred to the electronic version of the article.)}
\label{mans2}
\end{figure}

Before looking at numerical plots we have to consider three minor problems. First, it is too difficult to produce plots giving a good impression of these manifolds in the 5-dimensional energy shell of the phase space. Second and in addition, our further discussions will focus on the position space. These points are taken care of by projecting $W^u(S_E)$ into the position space. Third, the complete unstable manifold has an infinite extension and has an infinity of folds and tendrils and the outer branches leave to regions very far from the cluster. For our further discussion only the local segments are important, i.e. the parts emanating directly from the saddle region. In addition, we can expect that the tidal approximation no longer makes sense in regions far from the cluster. Therefore we cut off the unstable manifold, we restrict it by following the above mentioned orbits over a finite time only. With these considerations in mind we plotted in Fig. \ref{mans2} the projection into the $(x,y,z)$ position space of the local segments of the unstable manifolds of the saddle NHIMs. The time interval used for the cut off is [0, 4.5]. In this figure, stars in the inner region of the cluster are coloured blue. Orbits leaving the cluster over the saddle $L_1$ are plotted red, while orbits leaving over the saddle $L_2$
are plotted green.

This figure shows a very important property: The unstable manifolds of the saddle NHIMs and therefore also the tidal tails of the cluster stay in a rather thin layer around the plane $z = 0$. This means, that the tidal arms are aligned in the symmetry plane of the parent galaxy.

\begin{figure*}
\centering
\resizebox{\hsize}{!}{\includegraphics{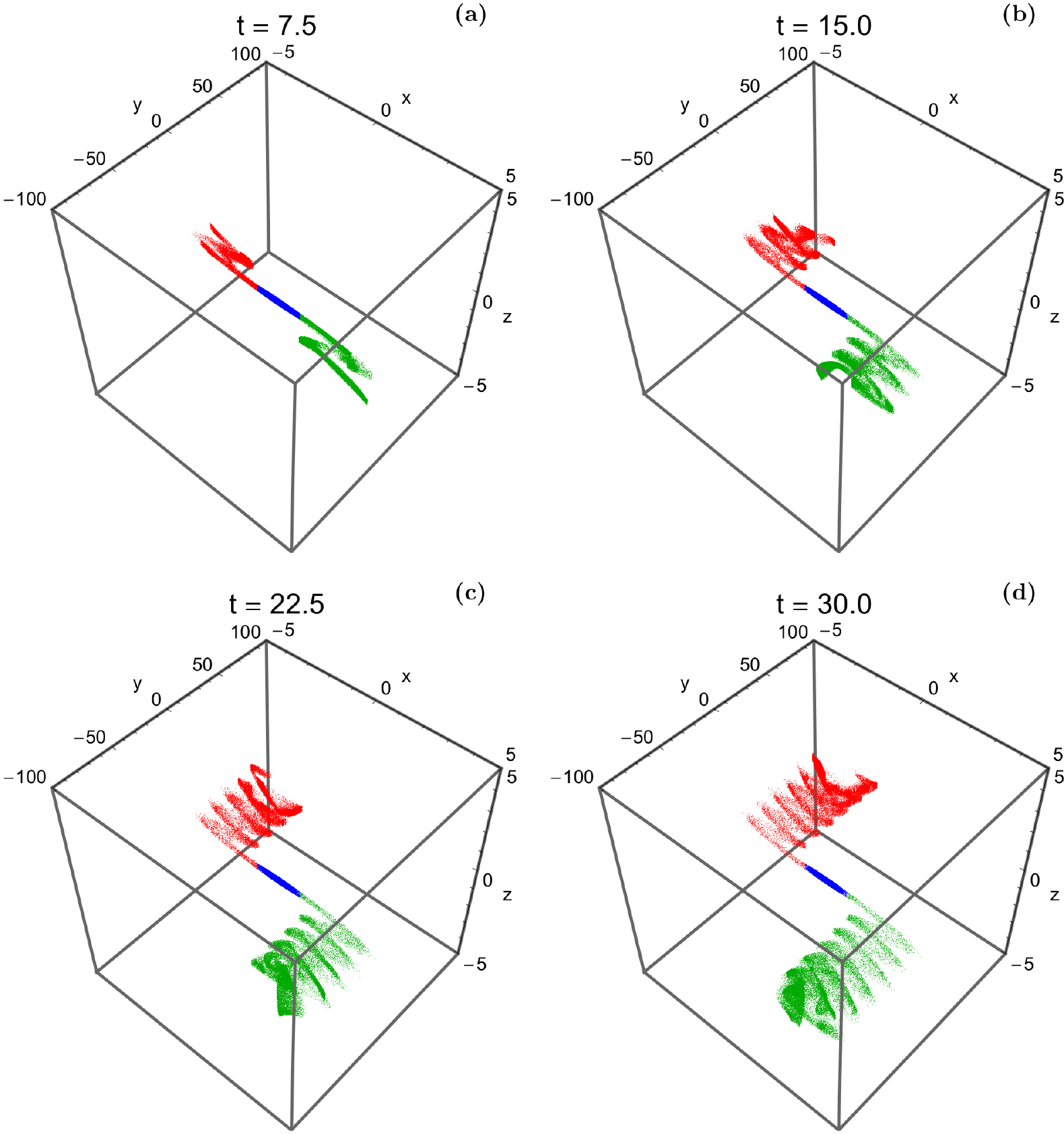}}
\caption{The distribution of the position of stars in the configuration $(x, y, z)$ space initiated $(t = 0)$ within the Lagrange radius, for $E = -3$. The leading tail (red) contains stars that escaped through $L_1$, while the trailing tail (green) contains stars that escaped through $L_2$. Stars moving in bounded (trapped chaotic or non-escaping regular) motion inside the Lagrange radius are shown in blue. Note that the development of the tidal tails takes place very close to the primary $(x,y)$ plane. (For the interpretation of references to colour in this figure caption and the corresponding text, the reader is referred to the electronic version of the article.)}
\label{tt}
\end{figure*}

Within the tidal approximation the two unstable manifolds, i.e. the one from saddle $L_1$ and the one from saddle $L_2$, have the same shapes and are transformed into each other by the inversion of horizontal coordinates. Without the tidal approximation, i.e. under the full potential of the parent galaxy, these two manifolds should develop different shapes on a large scale because they are developed under different influences from the parent galaxy.

For a better understanding of the behaviour of escaping stars, we reconfigured our numerical integration routine for the computation of the basins of escape to yield output of all orbits. For the initial condition of the orbits we define a dense
uniform three dimensional grid of size $N_x \times N_y \times N_z = 100 \times 100 \times 100$, with $p_{x0} = p_{z0} = 0$, while both branches of $p_{y0}$ (obtained through the Jacobi integral of motion) are allowed. All the initial conditions of the orbits lie inside the Lagrange radius of the star cluster, with $E = -3$. We numerically integrate the initial conditions of the three-dimensional orbits and we record the output of all orbits. This allow us to monitor the formation as well as the time-evolution of the tidal tails constructed by the stars that escape through $L_1$ and $L_2$.

Initially, at $t = 0$ all orbits are regularly distributed within the Lagrange radius. In Fig. \ref{tt}(a-d) we present the time-evolution of the $(x, y, z)$ position of the stars. It is seen that as time evolves both tidal tails (leading and trailing) grow along the $y$-axis however, the values of the $x$ and $z$ coordinates remain always very low ($|x| < 5$ pc and $|z| < 0.2$ pc ). The numerical integration of the orbits was carried out until $t_{max} = 30$ so that $y_{max} \simeq 100 \ r_{\rm L}$. We observe in panel (a) of Fig. \ref{tt} that at $t = 7.5$ time units the majority of stars are still inside the Lagrange radius however, a small portion of stars has already escaped passing through either $L_1$ or $L_2$. As time goes by we observe in panels (b-d) that the two tidal tails grow in size and at $t = 30$ time units the stellar structures are fully developed, while at the same time the area inside the tidal radius is uniformly filled. Here it should be pointed out that the standard integration code, which was used to obtain the results presented in Figs. \ref{xz1} and \ref{xz2}, records a point at every step of the numerical integration. In Fig. \ref{tt}(a-d) on the other hand, the density of points along one star orbit is taken to be proportional to the velocity of the star. Being more specific, a point is plotted (showing the position of a star), if an integer counter variable which is increased by one at every integration step, exceeds the velocity of the star. Following this technique we can simulate, in a way, a real $N$-body simulation of the tidal tails evolution of the star cluster, where the density of stars will be highest where the corresponding velocity is lowest. Additional numerical simulations, not presented here, reveal that the same behavior occurs for other (lower or higher) values of energy.

With a much closer look at Fig. \ref{tt}(a-d) it is easy for someone to observe that the two developed tidal tails are not symmetrical. Our calculations indicate that during the entire time interval of the development of the stellar structures the portion of stars that escape through $L_2$ is slightly elevated with respect to the amount of stars that escape through $L_1$. Being more precise, exit channel 2 appears to be about 6\% more preferable than exit channel 1. This observation comes in an agreement with the results presented earlier in Fig. \ref{percs}, where we seen that the percentage of escaping orbits through $L_2$ is always higher than that the rate of escaping orbits through $L_1$. Only when the energy level is relatively close to the critical energy of escape these two percentages coincide due to the complete fractality of the phase space. At this point, we would like to clarify that the deviation of the escape rates of the two channels is due to the particular choice of the initial conditions of the orbits.

\section{Conclusions}
\label{conc}

The aim of this work was to numerically investigate the 3-dof escape dynamics of a star cluster embedded in the steady tidal field of a parent galaxy. We considered energy levels larger than the escape energy where the equipotential surface opens and two exit channels appear, through which the test particles (stars) can escape to infinity. We managed to distinguish between ordered versus chaotic and trapped versus escaping orbits. We also located the basins of escape leading to different exit channels, finding at the same time correlations with the corresponding escape time of the orbits. Our extensive and thorough numerical exploration strongly suggests that the overall escape mechanism is a very complicated procedure.

The NHIMs over the saddle points act as the most important elements of the skeleton of the global dynamics as they direct the escape process over the saddles. Therefore we studied the development scenario as function of the total orbital energy for the NHIM over the saddle point $L_2$ in detail and this includes the development scenario of the Lyapunov orbits contained in this NHIM. The main tool to study the NHIM has been the numerical construction of the restricted Poincar\'e map on the NHIM.
It provides simultaneously the invariant surface itself and the dynamics on it.

The stable and the unstable manifolds of the NHIMs play an essential role because they transport important dynamical structures created on the NHIM to far away regions of the phase space and they channel and direct the escape flow over the saddles. As a consequence, the escaping orbits follow the unstable manifolds of the NHIMs and thereby the projection into the position space of these unstable manifolds delimits the tidal tails (or arms) of the cluster formed by the escaping stars.

The main numerical results of our research can be summarized as follows:
\begin{enumerate}
  \item Regions of bounded regular motion were found to coexist with extended basins of escape composed of initial conditions of orbits that escape through the exit channels located near the two saddle points.
  \item Bounded regular orbits mainly correspond to simple 1:1 loop three-dimensional orbits, while other secondary resonant orbits were also observed. The shift of the value of the total orbital energy mostly influences the orientation of these loop orbits. In particular, we found that at relatively high energy levels the 1:1 loop orbits become inclined to the $(x,y)$ plane.
  \item At energy levels above, yet very close to the critical energy of escape, we detected the existence of a substantial amount of chaotic orbits which remain trapped inside the Lagrange radius for vast time intervals. However the percentage of these trapped chaotic orbits heavily reduces as we proceed to higher energy levels.
  \item A strong correlation between the value of the Jacobi integral of motion and the extent of the basins of escape was found to exist. More precisely, for low energy levels the structure of the $(x,z)$ plane exhibits a large degree of fractalization which implies that the majority of the orbits escape choosing randomly escape channels. As the value of the energy increases the orbital structure becomes less and less fractal, while several basins of escape emerge and dominate.
  \item It was revealed that the percentages of the orbits that escape through the two escape channels are almost equal only very close to the energy of escape, where the degree of fractalization is still strong. For higher values of the energy on the other hand, their rates start to diverge and at relatively high energy levels channel 2 (around $L_2$) seems to be two times more preferable with respect to channel 1. However this behaviour is due to the particular choice of the initial conditions of the orbits.
  \item Our numerical analysis on the development scenario regarding the families of the periodic orbits contained in the NHIMs suggests that the orbital content of the network of the periodic orbits is much simpler with respect to other physical systems (i.e., a 3-dof model of a barred galaxy). In other words, our computations indicate that there are less bifurcated orbits of the main Lyapunov orbits.
  \item We provided numerical evidence that stars with initial conditions in the close vicinity of the NHIMs of the two Lagrange points $L_1$ and $L_2$ can form tidal tails (or tidal arms) upon escape through the exit channels around the saddles.
\end{enumerate}

We hope that the results of the present numerical exploration shed some light on the properties of the escape dynamics as well as on the role of the normally hyperbolic invariant manifolds (NHIMs) in tidally limited star clusters.

\section*{Acknowledgments}

One of the authors (CJ) thanks DGAPA for financial support under grant number IG-100616. We would like to express our warmest thanks to the anonymous referee for the careful reading of the manuscript and for all the apt suggestions and comments which allowed us to improve both the quality and the clarity of our paper.

\section*{Appendix A: Relation between the total and the effective potential}
\label{app1}

In previous works on the subject of tidally limited star clusters (e.g., Papers I and II) the effective potential
\begin{equation}
\Phi_{\rm eff}(x,y,z) = \Phi_{\rm cl}(x,y,z) + \frac{1}{2}\left(\kappa^2 - 4\Omega^2 \right) x^2 + \frac{1}{2}\nu^2 z^2, \label{eff}
\end{equation}
has been used. However in this paper we decided to use the total potential $\Phi_{\rm t}$ given in Eq. (\ref{vt}). This is necessary because we used the canonical form instead of the velocity form for the equations of motion and the variational equations. In the following, we shall justify our choice and we will also make clear the connection between them. For any system of coordinates in position space we can write the equations of motion either in velocity form or in the canonical form. The canonical form always has the advantage that the symplectic structure of Hamiltonian dynamics becomes evident. Then we can apply the whole mathematical machinery of symplectic differential geometry without unnecessary complications. For a mathematically oriented presentation of Hamiltonian dynamics see \citet{AM78}.

Hamiltonian systems have some extra structure compared to more general dynamical systems, it is the symplectic structure. However, this symplectic structure only becomes evident in canonical coordinates. These are coordinates where the symplectic differential form $\omega$ has the simple functional form according to the Darboux theorem $\omega = \sum\limits_{j = 1}^{N} dp_j \wedge dq_j$, where $j$ runs over all degrees of freedom. These coordinates are not unique. A transformation from one coordinate system with the Darboux form to another one which also has the Darboux form is called canonical transformation. Only in these coordinates the area and volume conservation in various dimensions of the flow and of the Poincar\'{e} map is obvious. It is trivial to show that the time evolution itself is a canonical transformation. And, important for the following, only in canonical coordinates the relation between the total potential $\Phi_{\rm t}$ and the effective potential $\Phi_{\rm eff}$ is easy to understand.

Let the Hamiltonian in the inertial frame have the standard form $H = T + \Phi_{\rm t}$, where $T$ is the usual kinetic energy and $\Phi_{\rm t}$ is the total potential as a function of position coordinates $q$ (actually q is a multi component). The angular momentum is the generating function for rotations. Here one important property of canonical coordinates comes in: Any quantity is the generating function for translations in the conjugate direction. Therefore the transition to a rotating coordinate system with rotation frequency $\Omega$ is just the coordinate system where $\phi$ moves with constant velocity and therefore it is just generated by $- \Omega L$, where $L$ is the angular momentum. Therefore in order to obtain the Hamiltonian in the rotating frame of reference we just add $- \Omega L$ to the Hamiltonian in the inertial frame. The effective potential $\Phi_{\rm eff}$ is defined as $\Phi_{\rm eff}(q) = \min_p H(q,p)$, where we have to take the minimum of $H$ over all possible values of $p$ for fixed $q$. When $H$ is a smooth function then the minimum is obtained at the point where $\partial H / \partial p = 0$. But this partial derivative is just the velocity according to the Hamiltonian equations of motion. In our particular case this condition of velocity 0 means that $p_x = - \Omega y$, $p_y = \Omega x$ and $p_z = 0$. Thus we obtain the effective potential when we plug in these values for the momenta into the Hamiltonian in the rotating frame of reference. In our case the result is $\Phi_{\rm eff} = \Phi_{\rm t} - \frac{\Omega^2}{2} \left(x^2 + y^2\right)$. More details regarding the general method for constructing effective potentials can be found in \citet{JT94,M92,S70a,S70b}.

Now let us imagine that we have a good approximation for $\Phi_{\rm eff}$, usually it is some truncation of a power series expansion. However in $H$ we need $\Phi_{\rm t}$. Therefore we invert the above relation and we write $\Phi_{\rm t} = \Phi_{\rm eff} + \frac{\Omega^2}{2} \left(x^2 + y^2\right)$ and finally we insert the approximation for $\Phi_{\rm eff}$ into the Hamiltonian. For the tidally limited cluster we are exactly in the same situation. The tidal approximation gives a simple functional form for $\Phi_{\rm eff}$ and we can apply all this machinery to the star cluster.

\section*{Appendix B: Loops of orbits around the NHIM}
\label{app2}

If an initial condition lies exactly on the stable manifold of some unstable (hyperbolic) invariant subset, then the corresponding orbit converges to this subset and remains on it for ever. Of course, numerically we can never start exactly on a stable manifold, we can only start close to it. Then the orbit first approaches the invariant subset, stays close to it for a finite time and later leaves the neighbourhood of this subset again near and along its unstable manifold. The time which the orbit spends in the neighbourhood of the invariant subset depends on how close to the stable manifold the orbit has been started. If we come closer to the stable manifold by a factor eigenvalue of the invariant subset, then the orbit makes one more loop close to the invariant subset.

For 2-dof systems the relevant subsets are unstable (hyperbolic) periodic orbits. And then coming closer to its stable manifold by one eigenvalue means making one additional revolution close to this periodic orbit. For an instructive numerical example of this behaviour see Fig. 7 in \citet{JS88}. Additional and more general explanations of the scaling behaviour of escaping orbits in the neighbourhood of stable manifolds of invariant subsets can be found in \citet{JP89}.

\begin{figure*}
\centering
\resizebox{\hsize}{!}{\includegraphics{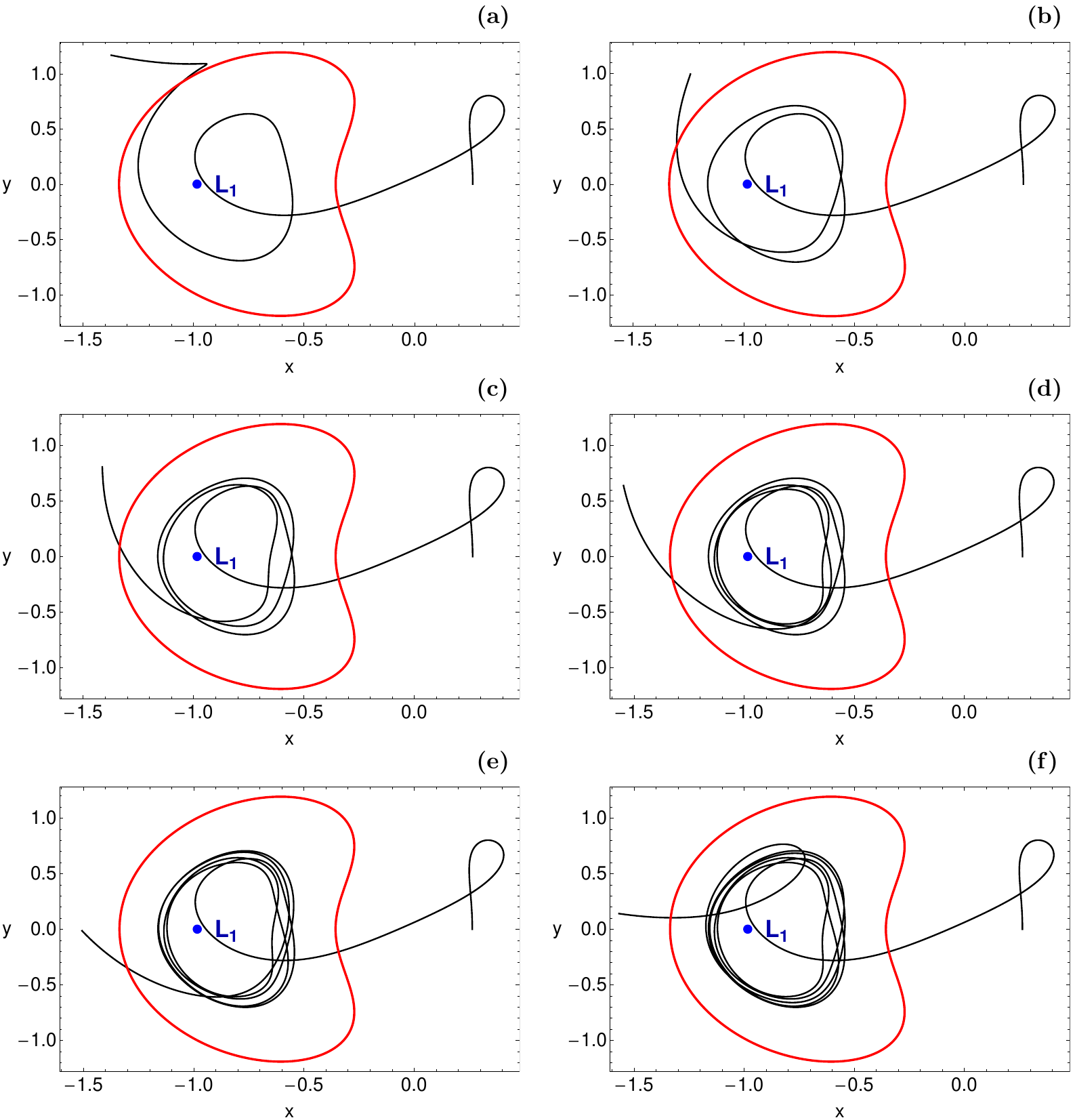}}
\caption{The projection on the $(x,y)$ plane of a three dimensional orbit with initial conditions: $y_0 = 0, z_0 = 0.3, p_{x0} = p_{z0} = 0$. The value of $p_{y0}$ is obtained from the energy integral with $E = -1.5$. The numerical value of the initial $x$ coordinate is $x_0 = 0.26427069432926$, while the number of the decimal digits of $x_0$ varies through the panels as follows: (a): four digits; (b): six digits; (c): eight digits; (d): ten digits; (e): twelve digits; (f): fourteen digits. The position of the Lagrange point $L_1$ is indicated by a blue dot, while the outermost red solid line is the horizontal Lyapunov orbit. We observe that the number of loops around $L_1$ increases with increasing accuracy of $x_0$. (For the interpretation of references to colour in this figure caption and the corresponding text, the reader is referred to the electronic version of the article.)}
\label{loops}
\end{figure*}

The above-mentioned early examples treated 2-dof system. However, the basic idea holds the same for any number of degrees of freedom. For NHIMs of codimension 2 the relevant eigenvalue is the one in normal direction. A numerical example from the present system of the star cluster is presented in Fig. \ref{loops}. From one part of the figure to the next one 2 more decimal digits have been included into the initial condition of the $x$ coordinate, And the sequence of initial conditions converges to the stable manifold of the NHIM. We clearly see how the number of loops which the orbit makes over the NHIM, and also around the Lagrange point $L_1$, increases with increasing precision of this initial condition.

\bsp
\label{lastpage}

\end{document}